\documentclass[preprint, aps, pre]{revtex4}
\usepackage{graphicx, bm, bbm,amsmath, amssymb, epsfig}

\begin{document}
\newcommand{\bq}{\begin{equation}}
\newcommand{\eq}{\end{equation}}
\newcommand{\nn}{\noindent}
\title{Bending of bilayers with general initial shapes}
\author{Silas Alben}
\affiliation{School of Mathematics, Georgia Institute of
Technology, Atlanta, GA 30332-0160} \email{alben@math.gatech.edu}
\date{\today}

\baselineskip=30pt
\begin{abstract}
We present a simple discrete formula for the elastic energy of a bilayer. The formula is convenient for
rapidly computing equilibrium configurations of actuated bilayers of general initial shapes.
We use maps of principal curvatures and minimum-curvature direction fields to analyze configurations.
We find good agreement between the computations and an approximate analytical solution for the case of a rectangular bilayer. For more general shapes (simple polyiamonds), we find a range of
typical bending behaviors:
overall bending directions along longest and shortest dimensions, inward bending at corners,
curvature intensification near boundaries, and conical bending and partitioned bending zones in some cases.
\end{abstract}
\pacs{46.40.Ff, 46.40.Jj, 47.15.Ki, 47.52.+j} \maketitle

\section{Introduction}

A bilayer is a thin sheet consisting of two layers, each composed of a different material. When the environment external to the bilayer undergoes a change in temperature or chemical composition,
the equilibrium strains of the two layers change by different amounts due to their different material
properties \cite{timoshenko1925analysis}. In one common type of bilayer \cite{smela1993electrochemical}, one layer, called the substrate, has an
equilibrium strain of zero throughout the external environmental change. The
other layer, called the actuated layer, has an equilibrium strain which changes from zero
to a (possibly nonuniform) value $\epsilon_a$. Because the layers are bonded,
they undergo the same strain at their interface, and thus both layers cannot
simultaneously be uniformly at their equilibrium strains. However, each layer can be brought
closer to its equilibrium strain, on average, when the bilayer bends. Then
the layer along the outer circumference is stretched relative to the inner layer, so
the average strain in each layer is different.

Bilayers are a widely-used technology for producing precisely-controllable motions and
shape changes for solid bodies. A common application is the household thermostat \cite{timoshenko1925analysis}, and newer
applications include self-assembling containers \cite{smela1995controlled,leong2008thin}, biomedical devices \cite{smela2003conjugated}, and
radio-frequency switches for wireless communications \cite{guerre2010wafer}.
Bilayers can be considered as a method for making new three-dimensional structures by inducing in-plane
stress. Recent work has used imposed stresses in deformable membranes to create wrinkled patterns \cite{sharon2007geometrically,huang2010smooth} and other buckled and twisted shapes \cite{klein2007shaping,armon2011geometry}.

In this study we extend a model for thin homogeneous elastic sheets to bilayers.
The previous model has been used to study a wide range of physical problems, including elastic strains due to defects in materials \cite{Seung:1988}, the crumpling of paper \cite{Lobkovsky:1997},
the self-assembly of elastic sheets under magnetic forces \cite{alben2007self}, and the bending and buckling of spheroidal pollen grains under osmotic pressure \cite{katifori2010foldable}. The simplicity of the model has made it very easy to apply to a wide range of problems. 

The present model represents the bending of a bilayer in terms of a mesh for just a single
surface---the central surface of the substrate layer. In terms of the shape of this surface,
we derive a discrete local formula for the strain in the central surface of the actuated layer.
Simple formulas for the elastic energy and its gradient are then obtained, which can be used
to find bilayer shapes as energy minima using standard optimization methods.
The computational cost of simulating a bilayer in this way is only slightly higher
than that of simulating a homogeneous elastic sheet, because the strain in the central surface of the
actuated layer is proportional to the curvature of the substrate.

The organization of the paper is as follows. In section \ref{Model},
we derive the bilayer model. In section \ref{Rect} we present sample simulations of
a rectangular bilayer, and compare with an approximate analytical solution. Section
\ref{General} presents examples of bilayer bending for various
bilayer shapes, most of which are simple polyiamonds. Section \ref{Conclusion}
summarizes the results.

\section{Model \label{Model}}

We represent a bilayer in discretized form using a triangular mesh. The mesh is
an equilateral triangular lattice in the undeformed state.
The substrate layer has stretching and bending energies which approximate those of a
uniform, isotropic elastic sheet. Both energies may be obtained by summing simple quantities over
the edges of the lattice \cite{Seung:1988}. The stretching energy is
\bq
E_s = \frac{1}{2} C_s \sum_{i,j} \left(|{\mathbf r}_i - {\mathbf r}_j| - d_{eq}\right)^2. \label{Es}
\eq
\nn Here $C_s$ is a stretching stiffness constant, $d_{eq}$ is the length of the edges
in the undeformed mesh, and the sum is over distinct nearest-neighbor
pairs of points. The bending energy is
\bq
E_b = C_b \sum_{\alpha,\beta} (1 - {\mathbf n}_\alpha \cdot {\mathbf n}_\beta). \label{Eb}
\eq
\nn $C_b$ is a bending stiffness constant, and the sum is over distinct nearest-neighbor pairs
of {\it triangles}, with the unit normal vectors to each triangle given by ${\mathbf n}_\alpha$ and ${\mathbf n}_\beta$.
In the undeformed state, the mesh lies in the $x$-$y$ plane, and all of the normal vectors are
$\mathbf{\hat{e}}_z$ (i.e. pointing upwards in the $z$ direction). The normal vector maintains the same
direction with respect to its triangle during deformations.

Seung and Nelson showed that when the stretching strain is small and the radii of curvature
of the sheet are large relative to $d_{eq}$, $E_s$ and $E_b$ converge to the stretching
and bending energies of a uniform isotropic elastic sheet with Poisson ratio $\nu$ = 1/3 \cite{landau1986te,Seung:1988}. If the (3D)
Young's modulus of the sheet is $E$ and its thickness is $h$, then $C_s$ and $C_b$ are
proportional to the 2D Young's modulus and the bending modulus of the sheet:
\bq
C_s = \frac{\sqrt{3}}{2} E h\,;\quad
C_b = \frac{2}{\sqrt{3}} \frac{E h^3}{12(1 - \nu^2)}. \label{CsCb}
\eq

The substrate and actuated layers are assumed to have uniform thicknesses
which are both equal to $h$, and the same stretching and bending stiffness constants, for simplicity.
In the initial configuration, the lower surface of the substrate layer lies in the plane $z = -h/2$ and the
upper surface lies in the plane $z = h/2$. The lower surface of the actuated layer
is bonded to the upper surface of the substrate layer at $z = h/2$,
and the upper surface of the actuated layer lies in the $z = 3h/2$ plane. Throughout
the bilayer in its initial, flat configuration, there are
planar and parallel surfaces of material, lying in the planes of constant $z$ between $-h/2$ and $3h/2$. As the bilayer bends, these surfaces of material are no longer planar, but {\it are}
assumed to remain parallel. This assumption underlies the classical theories of the
bending of thin plates and bilayers (including the continuum versions of
(\ref{Es}) and (\ref{Eb})),
and allows the configuration of the
bilayer to be expressed entirely
in terms of the configuration of the central surface of the substrate (i.e.
the material initially lying in the plane $z = 0$). The central surfaces
of the substrate and actuated layers may be assumed to remain parallel
under large deformations, as long as the thickness of the bilayer is small
compared to its radii of curvature at all points, in all directions.

The aforementioned triangular mesh represents the central surface of the substrate (i.e.
the material initially lying in the plane $z = 0$). The central surface of the
actuated layer is parallel to that of the substrate, and displaced from it
by a distance $h$ in the direction of the substrate surface normal $\mathbf n$.

The actuated layer is actuated by setting its equilibrium stretching strain (uniformly here) to  $\epsilon_a$, while the equilibrium strain remains zero for the substrate. Bending causes different stretching strains to occur
in the central surfaces of the substrate and actuated layers, and thereby allows each surface to move
closer to its own equilibrium stretching strain. If the substrate central
surface has curvature $\kappa$ in a certain direction at a point, then the strain in the central surface of the
actuated layer in that direction is that at the corresponding point in the substrate plus $h\kappa$ (see appendix \ref{hkappa}). We consider the central surface of the actuated layer to be discretized by a triangular
mesh of points, which are those in the substrate mesh plus $h\mathbf n$. Then the strain in the actuated
layer mesh can be written in terms of that in the substrate mesh. Hence only the substrate mesh is required
to characterize the full bilayer geometry. The lengths of edges in the actuated layer mesh are those of
the corresponding edges in the substrate mesh, plus the stretching induced
by bending in the direction of the edge. This stretching is a discrete
approximation of the stretching strain $h\kappa$, times the edge length. In appendix \ref{dischkappa}
we derive a discrete formula for the edge stretching by using a locally-quadratic approximation for the bilayer
surface centered at midpoint of each edge. The formula is:
\bq
\mbox{Stretching due to bending} \approx \frac{h\sqrt{3}}{4} \sum_{k = 1}^4 \varphi_k, \label{dhkappa}
\eq
\nn where the angles $\varphi_k$ are those between the normals of the four triangles
adjacent to the two triangles which share the edge whose stretching is given.
These angles are shown schematically in figure \ref{fig:PolymerMeshSchematic2}. The formula for the stretching
energy of the actuated layer is then:
\bq
E_{s,a} = \frac{1}{2} C_s \sum_{i,j} \left(|{\mathbf r}_i - {\mathbf r}_j| +
\frac{h\sqrt{3}}{4}\sum_{k = 1}^4 \varphi_k - d_{eq} - \epsilon_a d_{eq}\right)^2. \label{Esa}
\eq
\nn Equation (\ref{Esa}) is similar to (\ref{Es}), but includes the additional stretching
due to bending in the actuated layer (\ref{dhkappa}), and the equilibrium stretching in
the actuated layer $\epsilon_a d_{eq}$. In the outer sum in $E_{s,a}$, we omit
the contribution from edges which are close to the boundary of the mesh, for which not
all four of the angles \{$\varphi_k$\} are defined. An alternative would be to use
one-sided approximations for the stretching
due to bending near the boundary, which may give a higher order of accuracy, at the expense of
increased complexity of the scheme. To fully realize the increased accuracy in arbitrary initial
planar geometries, an unstructured triangular mesh may be preferred. However, one of the
main benefits of our approach is its simplicity.

The bending energy of the actuated layer is close to that of the substrate layer.
In appendix \ref{hkappa} we show that the curvature $\kappa_a$ in
a given direction along the central surface of the actuated layer is the same
as the curvature of the substrate (at the corresponding point and same direction),
up to a relative error of $h\kappa$. Since $h\kappa \ll 1$ by assumption in our large deformation
plate theory, we take the curvature of the actuated layer to be
that of the substrate. The bending energy of the actuated layer is then the same as (\ref{Eb}).

The total elastic energy of our model bilayer is then
\bq
E_T = E_s + E_{s,a} + 2E_b. \label{ET}
\eq

\section{Rectangular bilayers \label{Rect}}

\begin{figure}
  \centerline{\includegraphics[width=5.8in] 
  {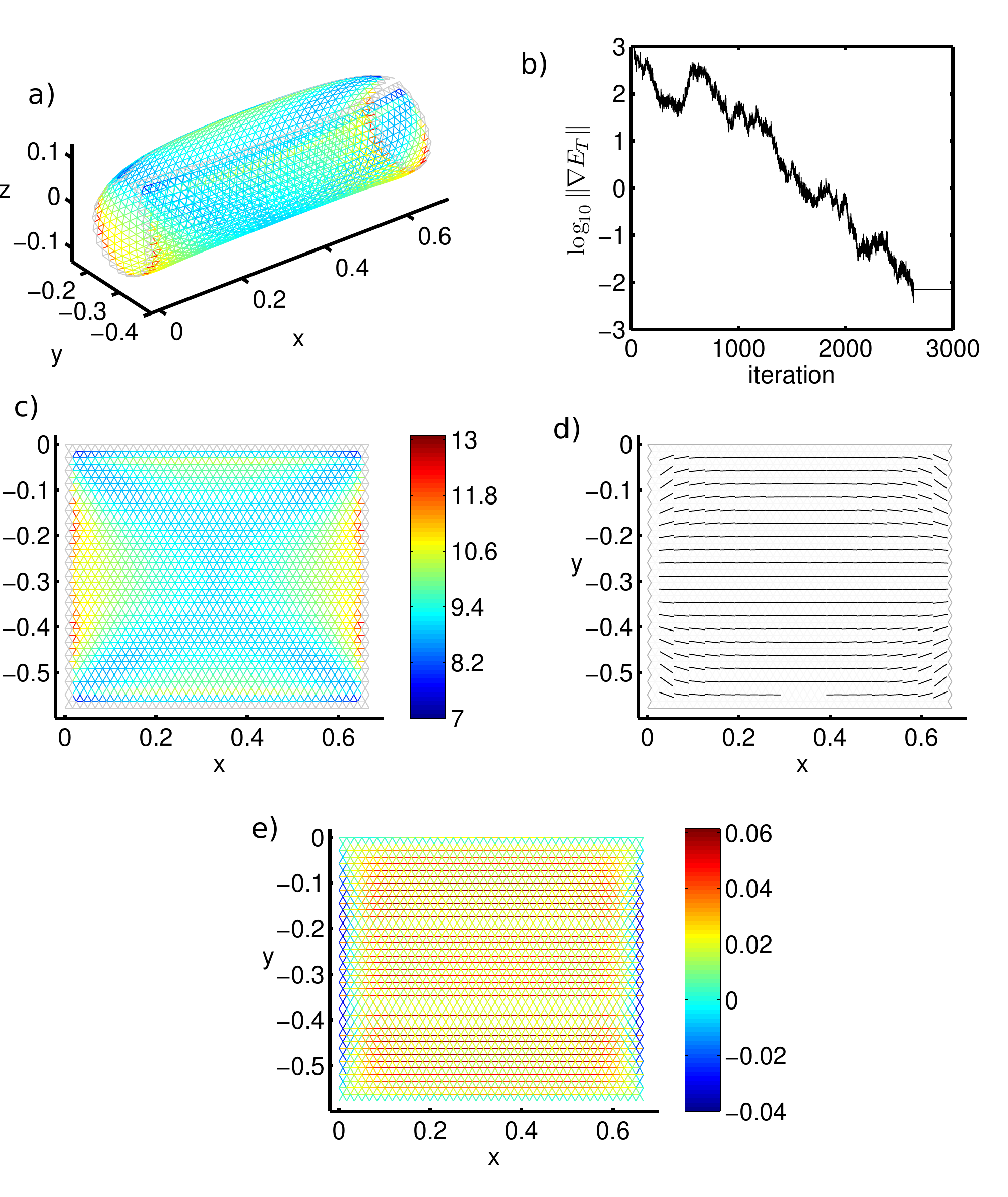}}
\vspace{-0.2in}
  \caption{Bending of a rectangular bilayer discretized
  with 1661 points and 4820 edges. The stretching stiffness
$C_s$ is $8\times10^4$, and the bending stiffness $C_b$ is 1. The thickness
is $h$ = 0.01, the initial height (along $y$) is $\sqrt{3}/3$, the initial width (along $x$) is $2/3$. The actuation strain $\epsilon_a$ is 10$h$.
a, Configuration of rectangular sheet in equilibrium, with bending along the shorter
dimension. b, 2-norm of energy gradient versus iteration number in LM-BFGS method, leading up to stagnation, when the equilibrium is reached. c, Color map of maximum principal curvature at the midpoints of edges. Gray edges are too close to the boundary to obtain estimates. d, Direction field showing the directions of minimum
principal curvature at the midpoints of edges. e, Color map of stretching
strain in each edge of the substrate central surface.
  }
  \label{fig:Rect1Fig}
\end{figure}

We now test our scheme for perhaps the simplest bilayers---rectangular bilayers---which have been studied previously \cite{timoshenko1925analysis,christophersen2006characterization,alben2011edge}.
We first consider a rectangular bilayer which is nearly square---aspect ratio $= \sqrt{3}/2 = 0.866$. The stretching stiffness
$C_s$ is $8\times 10^4$, and the bending stiffness $C_b$ is 1. By
(\ref{CsCb}), we then have that the thickness of each layer $h$ is 0.01. The height and width
are of order one ($\sqrt{3}/3$ and $2/3$ respectively),
so we have a thin sheet. We set
$\epsilon_a$ to $10 h$. At equilibrium, $h\kappa$ is the strain in the
actuated layer due to bending in the substrate, and this should be comparable
to $\epsilon_a$. Thus we expect $\kappa$ of about 10.

We use a mesh with equilibrium edge length $1/60$, yielding a discretization
with 1661 points (4983 degrees of freedom) and 4820 edges. We use a large-scale quasi-Newton scheme to minimize the discrete energy
(\ref{ET}), the limited-memory Broyden-Fletcher-Goldfarb-Shanno (LM-BFGS) method
with a cubic line search. We use an exact formula for the gradient of $E_T$ which can be computed quickly. An exact gradient formula handles large differences in elastic constants better than
a finite difference approximation. The initial guess for the minimization routine is the flat state.
In figure \ref{fig:Rect1Fig}a we show the equilibrium reached by the
minimization scheme after it stagnates at about 2500 iterations. The sheet has curled along its
shorter dimension, yielding a ``cigar'' shape, one of two equilibria previously found for
rectangular bilayers \cite{alben2011edge}. The decrease of
the 2-norm of $\nabla E_T$ with iteration number is shown in panel b, and
at stagnation, no further decrease of $E_T$ is obtained along the steepest
descent direction $-\nabla E_T$. At stagnation, the 2-norm of $\nabla E_T$
is $7\times10^{-3}$, and the $\infty$-norm is $4 \times 10^{-4}$.

Panel c gives a color map showing the larger of the two principal curvatures at the midpoints of the edges in the
equilibrium configuration in `a'. Here the bilayer is shown in its initial flat state for clarity.
The colors of the edges are the same in `a' and `c', so the correspondence between points on the initial
and final shape can be seen.
We estimate the principal curvatures using the method described in appendix \ref{pkappas}.
The estimate can be obtained for all edges except those adjacent to the boundary. Over most
of the bilayer the maximum principal curvature lies between 8.5 and 10.5. Near the boundary, there is more variation.
Panel d gives a field of line segments showing the directions of minimum-magnitude principal curvature at the edge midpoints. For a developable surface, the minimum-magnitude principal curvature is zero at each point. The zero-curvature direction field has integral curves that are straight lines,
called generators, which cover the
surface \cite{struik1988lectures}. The shape in panel a (and the other equilbrium shapes in this work) are approximately developable, so the lines in panel d approximate the generators of a developable surface. The lines are mostly horizontal, showing a cylindrical shape, except near the corners, which curl inward.

For cylindrical bending, we can compare our results with an approximate analytical continuum solution. Let us
assume that the bilayer bends with uniform curvature $\kappa$ in the $y$ direction,
and zero curvature in $x$.
Let us also assume uniform strains $\epsilon_x$ and $\epsilon_y$ in the $x$ and $y$ directions.
Then, by \cite{landau1986te,Seung:1988} the elastic energy per unit area over the bilayer
is uniform and equal to
\begin{align}
e_T &= \frac{\sqrt{3}}{8}C_s\left(3\epsilon_x^2 + 3\epsilon_y^2 + 2\epsilon_x\epsilon_y \right.\nonumber\\
&\left.+3(\epsilon_x-\epsilon_a)^2 + 3(\epsilon_y + h\kappa - \epsilon_a)^2 + 2(\epsilon_x-\epsilon_a)(\epsilon_y + h\kappa - \epsilon_a)\right)+\frac{\sqrt{3}}{2}C_b\kappa^2. \label{eT}
\end{align}
\nn In (\ref{eT}), the first three of the six terms in the sum multiplying $C_s$ correspond to
 the substrate stretching energy, and the second three correspond to the actuated layer stretching
energy. The last term corresponds to the combined bending energy. The equilibrium, found
by minimizing $e_T$ with respect to $\epsilon_x$, $\epsilon_y$, and $\kappa$, is
\bq
\epsilon_x = \frac{\epsilon_a}{2}; \quad \epsilon_y = 0;\quad
\kappa = \frac{\epsilon_a}{h}.
\eq
\nn For the parameters in this simulation,
\bq
\epsilon_x = 0.05; \quad \epsilon_y = 0;\quad
\kappa = 10.  \label{est}
\eq
\nn In figure \ref{fig:Rect1Fig}c, most of the curvature values (at edges away from the boundaries) lie between
8.5 and 10.5, close to the value of $\kappa$ in (\ref{est}). The curvature is somewhat
nonuniform in the simulation, unlike in the analytical approximation. Panel e
shows the stretching strain in each of the edges.
For the horizontal edges away from the sheet boundaries, the strain lies between 0.045 and 0.055,
with an average value close to $\epsilon_x$ in (\ref{est}). For the other edges (at $\pm \pi/3$ radians from
horizontal), the strain lies in the range 0.012--0.016, while that predicted by the analytical solution
is $\epsilon_x\cos^2\pi/3 + \epsilon_y\sin^2\pi/3 = 0.0125$.

\begin{figure}
  \centerline{\includegraphics[width=5.8in] 
  {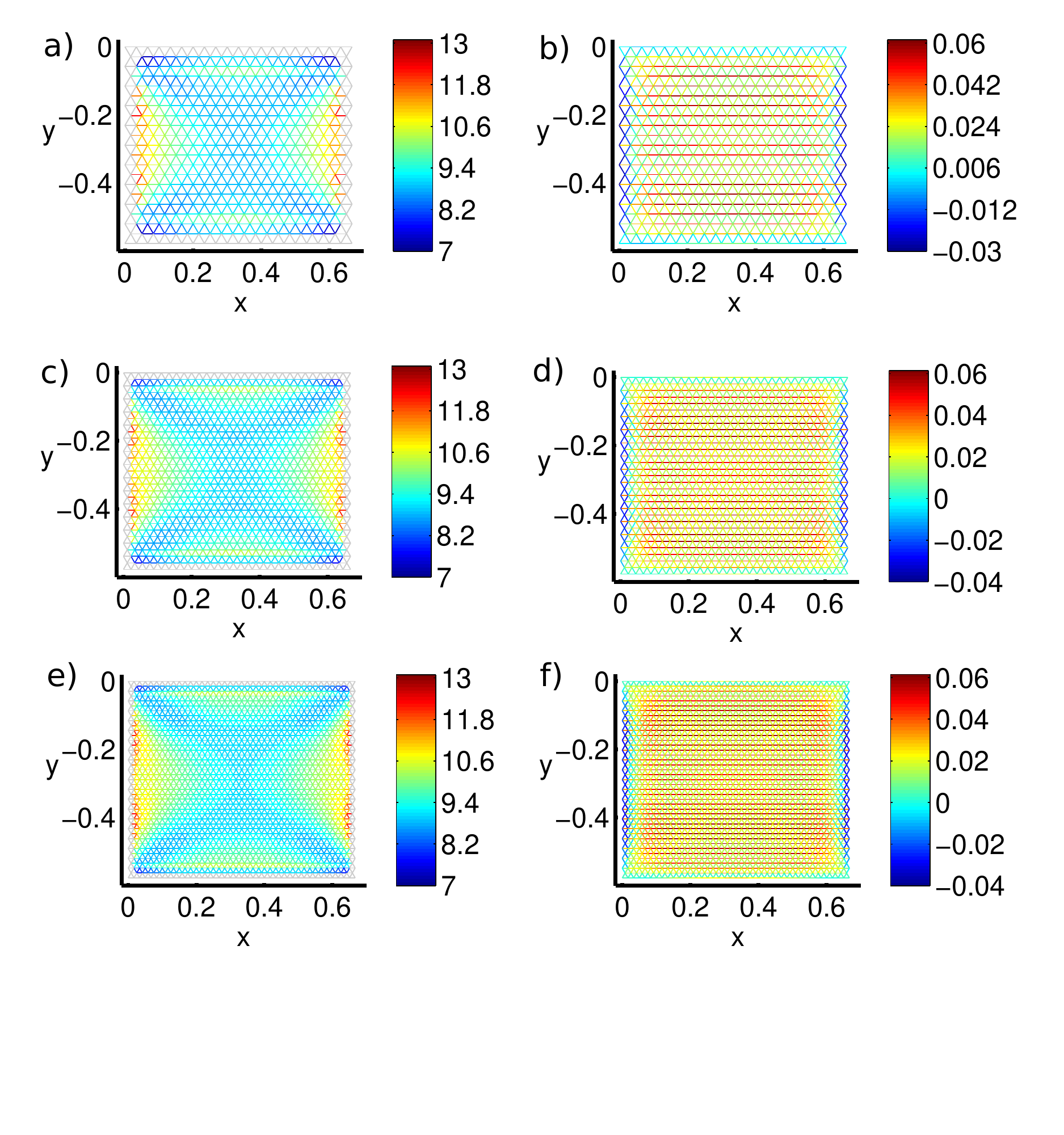}}
\vspace{-0.5in}
  \caption{Rectangular bilayer bending with the same parameters as in
figure \ref{fig:Rect1Fig} and three values of equilibrium
edge length: 1/30 (a,b), 1/45 (c,d) and 1/60 (e,f). Panels a, c, and e show the maximum principal curvatures
at the edge midpoints, and panels b, d, and f show the stretching strains in each edge.
}
  \label{fig:Rect2Fig}
\end{figure}

In figure \ref{fig:Rect2Fig}, we show the results from figure \ref{fig:Rect1Fig}c and e together with
solutions on two coarser meshes, with equilibrium edge lengths of 1/30 (a,b),
1/45 (c,d), and 1/60 (e,f). Panels a, c, and e compare maximum principal curvatures using the same color scale.
Panels b, d, and f compare stretching strains (with nearly the same color scales). The
distributions and overall magnitudes of these quantities are quite similar from the
finest to the coarsest mesh.

 \begin{figure}
  \centerline{\includegraphics[width=5.8in] 
  {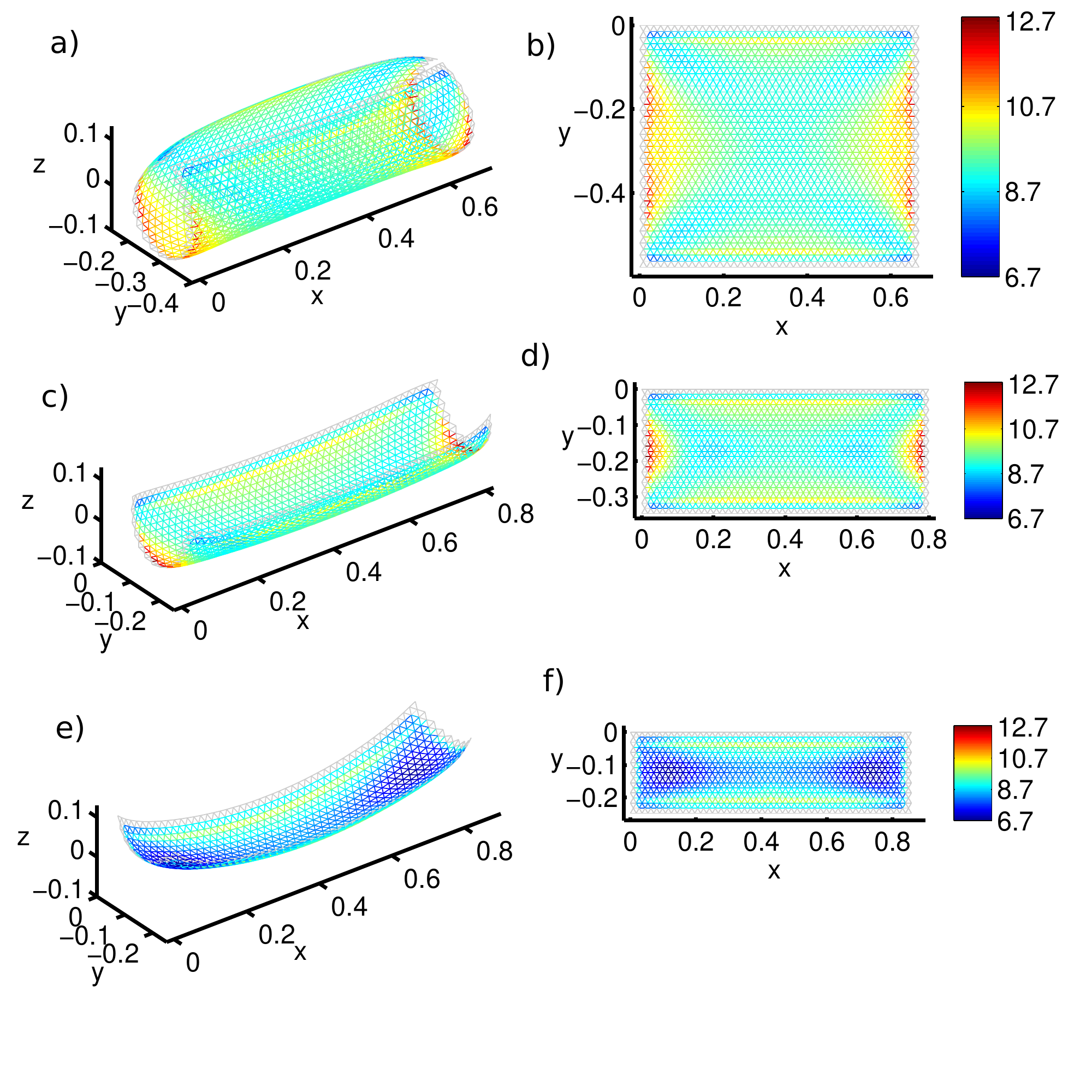}}
\vspace{-0.2in}
  \caption{Rectangular bilayer bending with the same parameters as in
figure \ref{fig:Rect1Fig} and three initial aspect ratios: (a) Aspect ratio = 1.15 (initial width = 2/3, initial
height = $\sqrt{3}/3 = 0.577$),
(b) Aspect ratio = 2.31 (initial width = 0.8, initial height = $\sqrt{3}/5 = 0.346$) (c) Aspect ratio = 3.46 (initial width = 6/7, initial height = $\sqrt{3}/7 = 0.247$).
}
  \label{fig:Rect3Fig}
\end{figure}

In figure \ref{fig:Rect3Fig}, we compare results across three aspect ratios of the initial rectangular
shape. All three shapes have maximum curvature along the short direction, although for the largest aspect
ratio (panel c) this is difficult to discern since the strip is much shorter in this direction, so it
does not show as much overall difference in the $z$ coordinate along this direction. Also, it has larger curvature in the long direction than the shapes in panels a and b. However, for
this strip the average curvature along the short direction is 10 times that in the longer direction.
The average of the curvature in the short direction decreases from about 9 for the most square
shape (a) to about 8 for the most elongated shape (c).

\section{General bilayers \label{General}}

We now consider the bending of a set of bilayers with more general planar shapes. Most of
the shapes we consider are connected clusters of a small number of equilateral triangles,
also called ``polyiamonds'', a class of polyforms \cite{weisstein1999crc}. Such shapes
can be represented with equilateral triangular meshes of various levels of refinement,
without jagged edges (as occur for the rectangles in figure \ref{fig:Rect3Fig}, for
example).

\begin{figure}
  \centerline{\includegraphics[width=5.8in] 
  {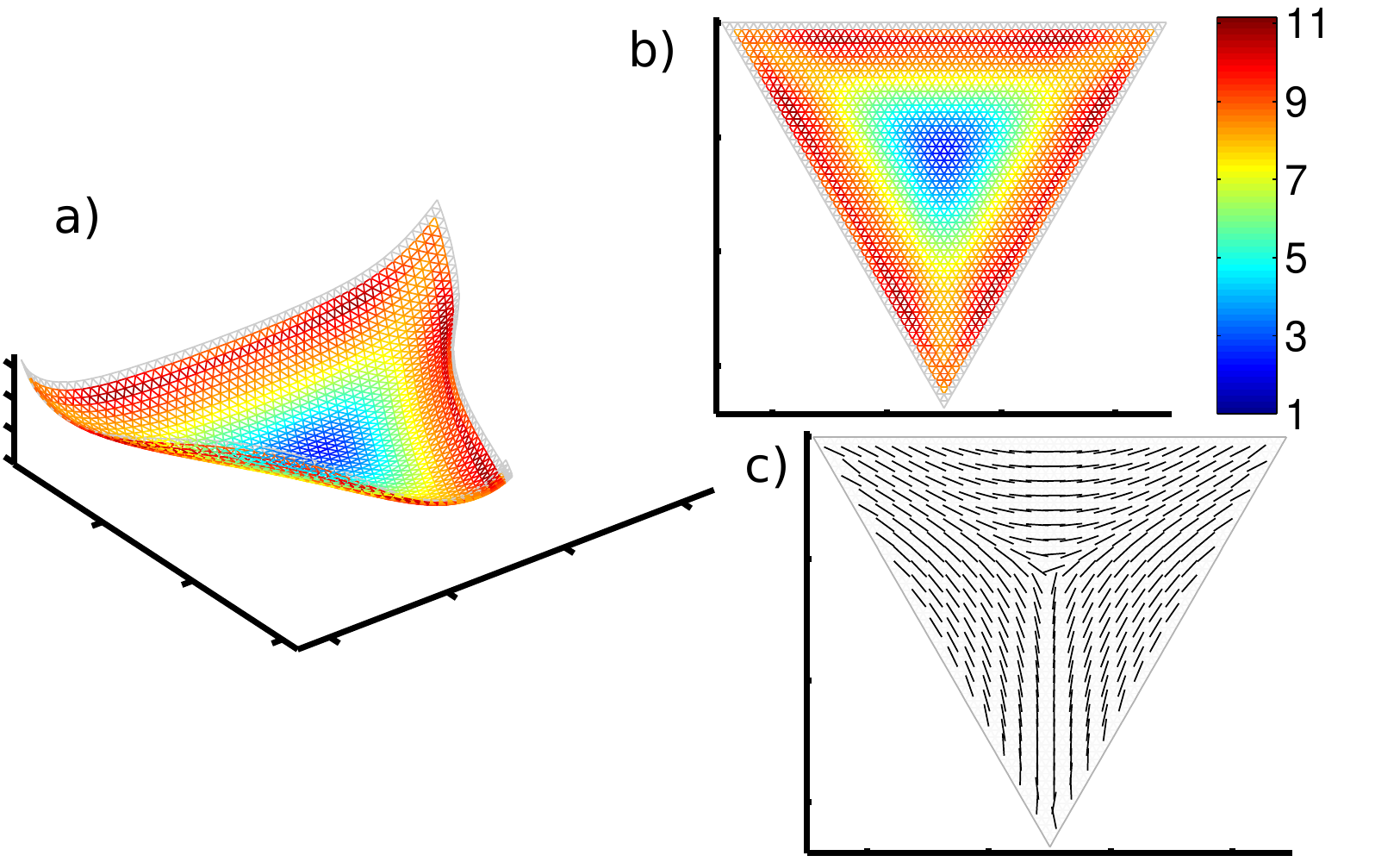}}
\vspace{-0.2in}
  \caption{Bending of an equilateral triangular bilayer with the same elastic parameters as in
figure \ref{fig:Rect1Fig}. The top edge has length 7/9. The values on the color bar indicate the maximum of the
principal curvatures at each edge midpoint. a, 3D configuration with coloring given in
b, Color map of maximum principal curvatures at each edge midpoint. c,
Direction field showing direction of minimum principal curvature at the midpoints of edges.}
  \label{fig:97}
\end{figure}

In figure \ref{fig:97} we show the bending of an equilaterial triangular bilayer with
the same elastic parameters as the rectangular bilayer just considered. The maximum curvature,
which occurs along the edges, is comparable in magnitude to the maximum curvatures of the
rectangles in figure \ref{fig:Rect3Fig}, but unlike for the rectangles, here there is a
large flattened region at the center. The minimum-curvature direction field (panel c) shows the triangular symmetry of
the shape. This equilibrum is not the shape with minimum elastic energy (or global equilibrium),
however. Cylindrical bending of the triangle is also an equilbrium, and has smaller total elastic energy.

\begin{figure}
  \centerline{\includegraphics[width=5.8in] 
  {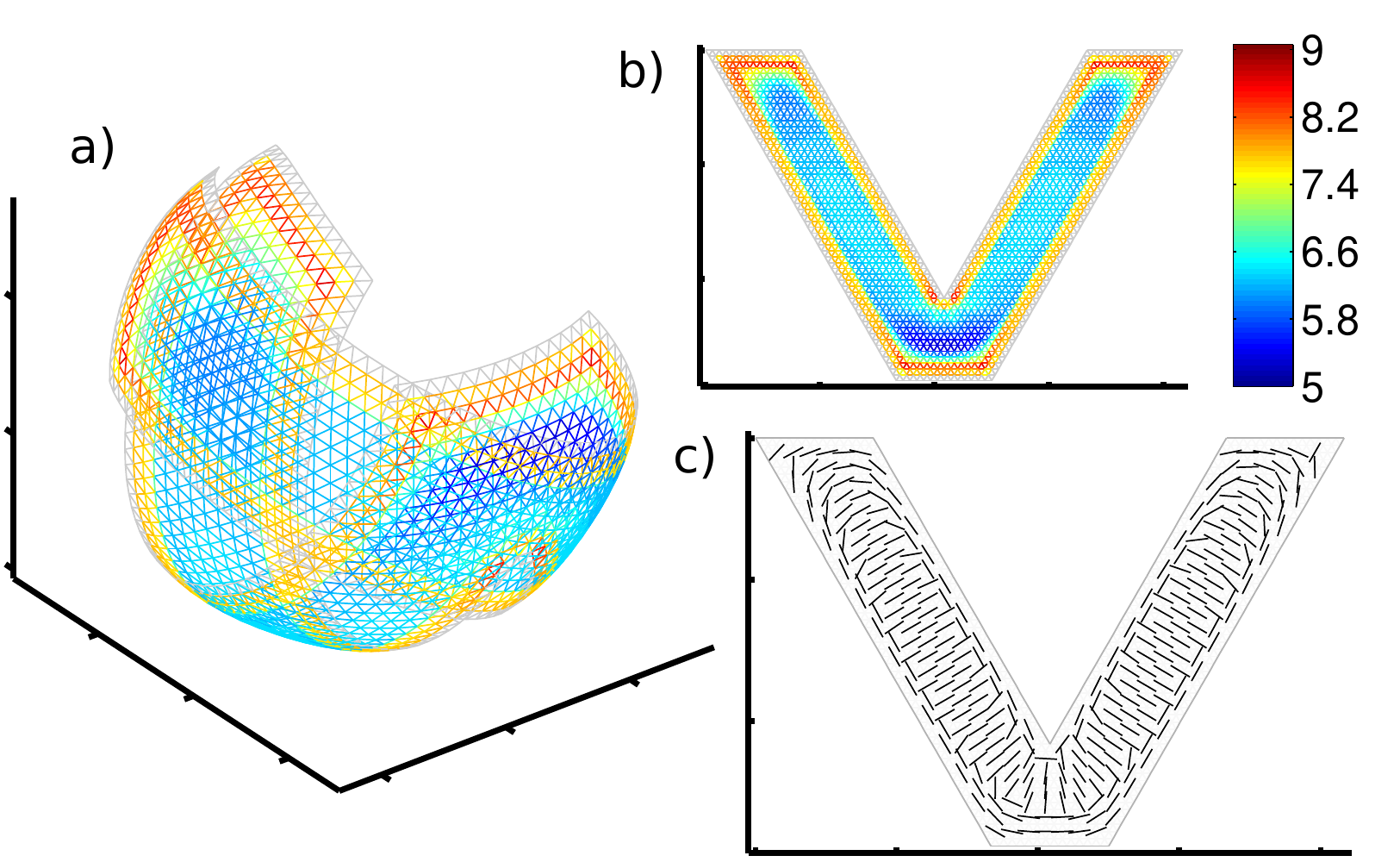}}
\vspace{-0.2in}
  \caption{Bending of a V-shaped bilayer with the same elastic parameters as in
figure \ref{fig:Rect1Fig}. The bottom edge has length 7/9. The values on the color bar indicate the maximum of the
principal curvatures at each edge midpoint. a, 3D configuration with coloring given in
b, Color map of maximum principal curvatures at each edge midpoint. c,
Direction field showing the directions of minimum principal curvature at the midpoints of edges.
}
  \label{fig:98}
\end{figure}

Figure \ref{fig:98} shows the bending of a V-shaped bilayer. Here the two arms of the V bend mainly along
the longer dimension, in contrast to the rectangular shapes in figure \ref{fig:Rect3Fig}, which bend along
the shorter dimension. Bending along the longer dimension is also an equilibrium for a rectangle,
with slightly lower elastic energy than for bending along the shorter dimension \cite{alben2011edge}.
For the V shape, there is a sharp transition to a region of higher curvature near the boundary, with
bending in the opposite direction. The minimum-curvature direction field in panel c shows that 
near the boundary, there is sharp transition in bending direction. Bending is nearly transverse
to the boundary near the edges, and parallel to the boundary further inside.

\begin{figure}
  \centerline{\includegraphics[width=5.8in] 
  {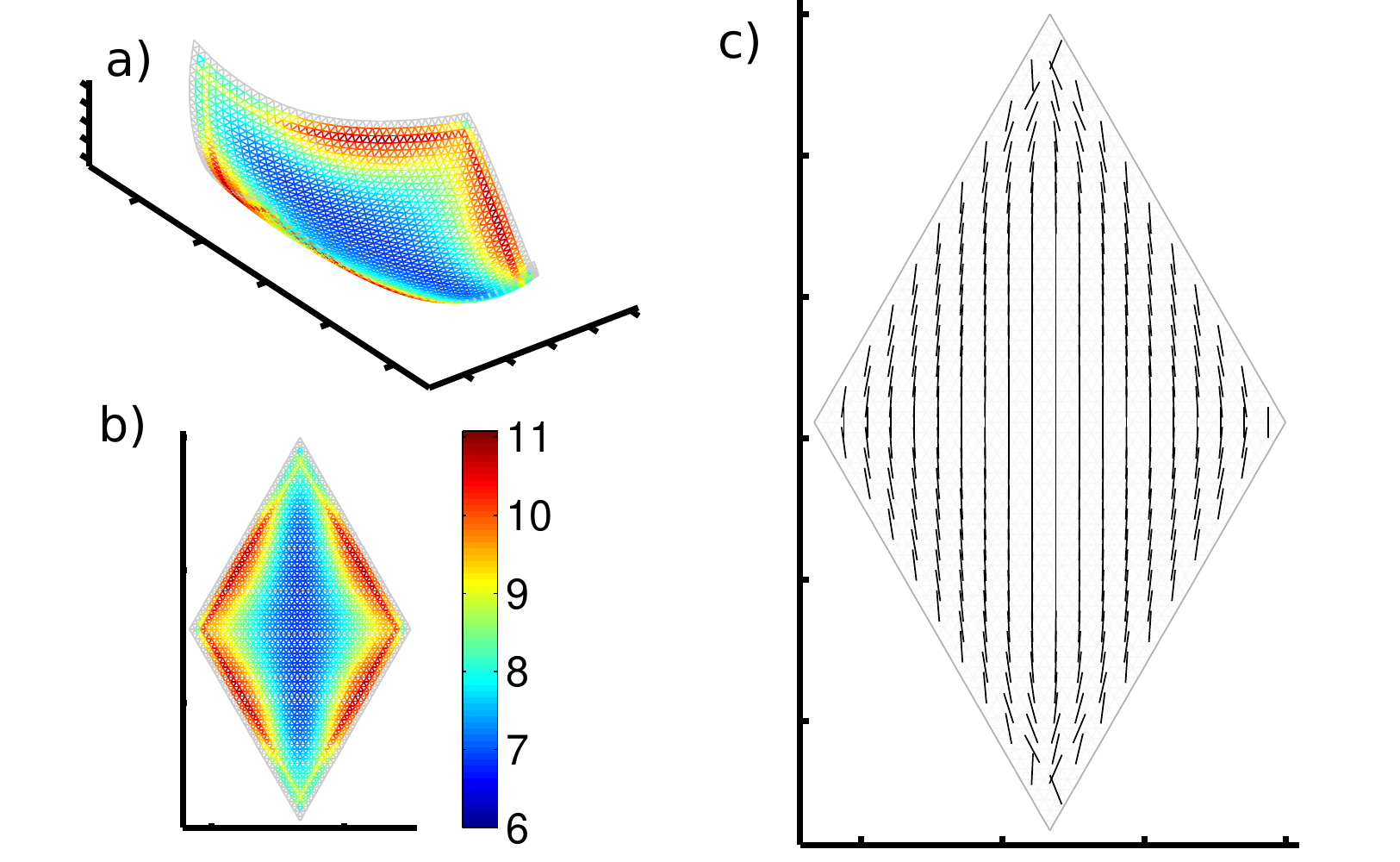}}
\vspace{-0.2in}
  \caption{Bending of a bilayer in the shape of a rhombus with the same elastic parameters as in
figure \ref{fig:Rect1Fig}. The horizontal width is 1/3. The values on the color bar indicate the maximum of the principal curvatures at each edge midpoint. a, 3D configuration with coloring given in
b, Color map of maximum principal curvatures at each edge midpoint. c,
Direction field showing the directions of minimum principal curvature at the midpoints of edges.
}
  \label{fig:100}
\end{figure}

Figure \ref{fig:100} shows the bending of a rhombus. Here the bending is mainly along the shorter dimension, akin to the rectangles of figure \ref{fig:Rect3Fig}. The direction of
bending does not change except at the farther pair of opposing tips. Near the edges there
are regions of increased curvature, larger than those in figure \ref{fig:Rect3Fig}.

\begin{figure}
  \centerline{\includegraphics[width=5.8in] 
  {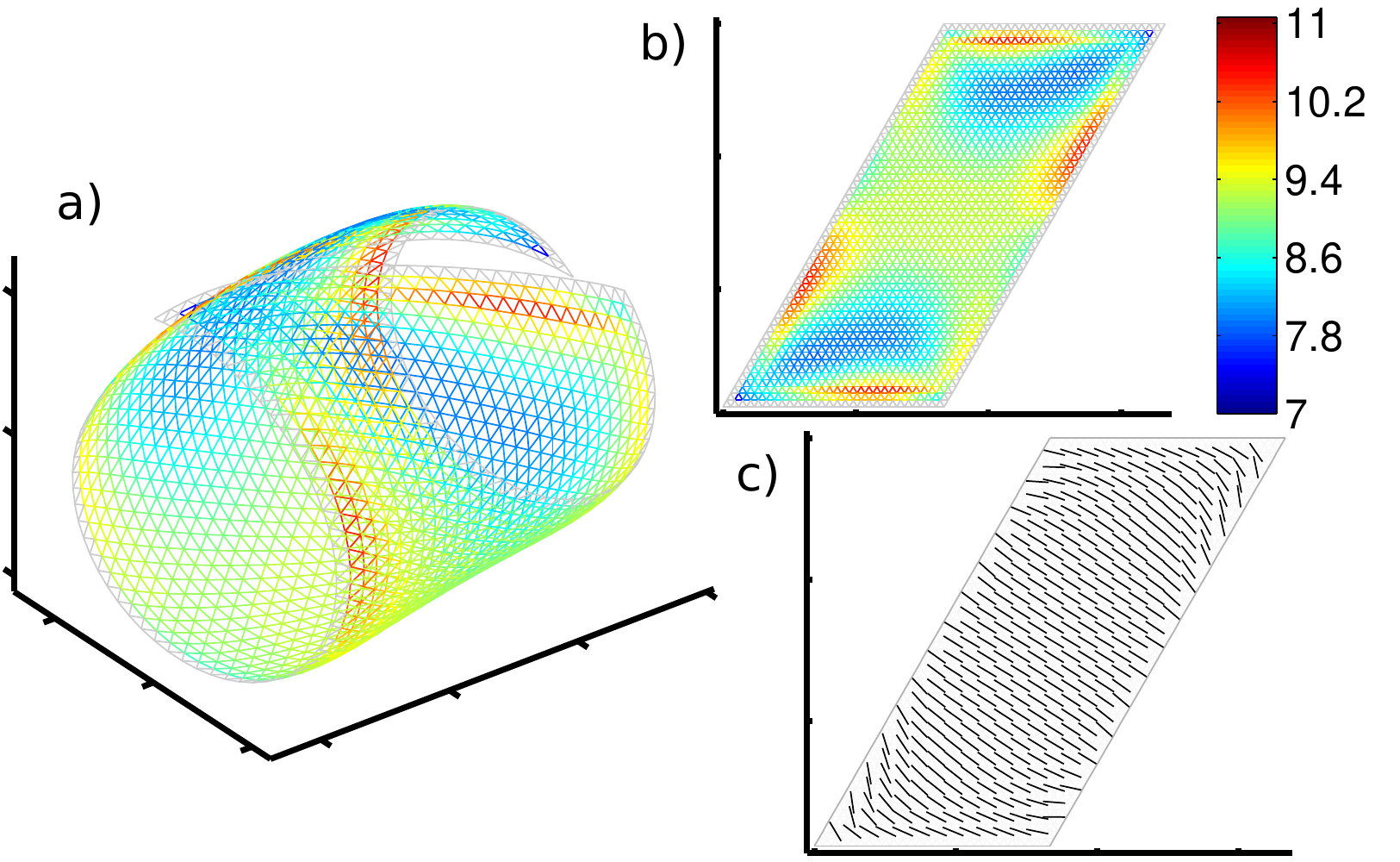}}
\vspace{-0.2in}
  \caption{Bending of a tetriamond bilayer with the same elastic parameters as in
figure \ref{fig:Rect1Fig}. The horizontal width is 1/3. The values on the color bar indicate the maximum of the principal curvatures at each edge midpoint. a, 3D configuration with coloring given in
b, Color map of maximum principal curvatures at each edge midpoint. c,
Direction field showing the directions of minimum principal curvature at the midpoints of edges.
}
  \label{fig:101}
\end{figure}

An oblique strip, a cluster of four equilateral triangles (a tetriamond), is shown in
figure \ref{fig:101}. The bending is essentially cylindrical, and this time along the longer
dimension. The minimum-curvature direction field is almost orthogonal to the longer edges, but there is a noticeable
deviation from orthogonality. The curvature is nearly uniform over the middle part of the
bilayer, and varies more near the tips with acute angles. The final shape maintains the 180-degree rotational symmetry of the initial shape.

\begin{figure}
  \centerline{\includegraphics[width=5.8in] 
  {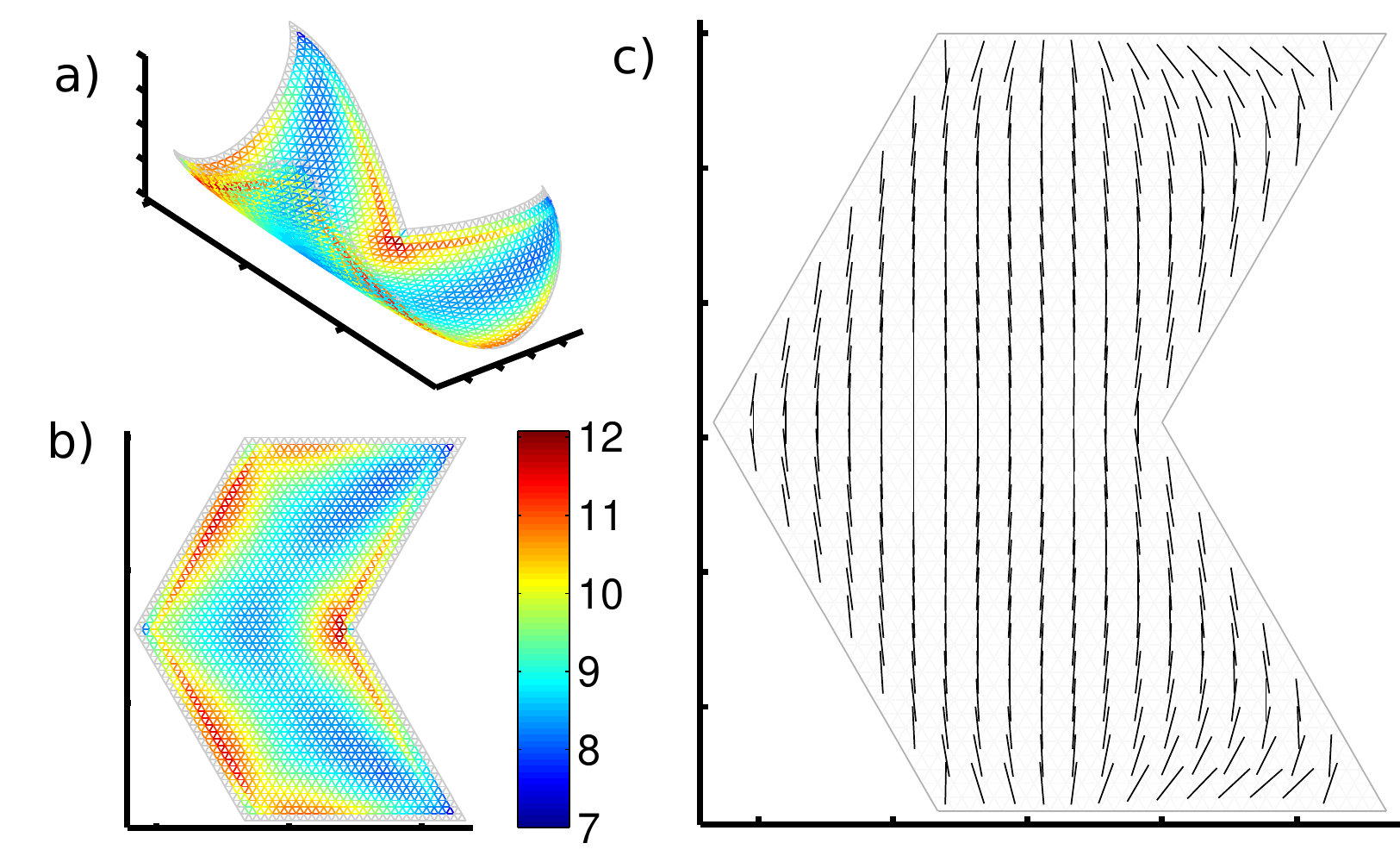}}
\vspace{-0.2in}
  \caption{Bending of a tetriamond bilayer with the same elastic parameters as in
figure \ref{fig:Rect1Fig}. The horizontal width is 1/3. The values on the color bar indicate the maximum of the principal curvatures at each edge midpoint. a, 3D configuration with coloring given in
b, Color map of maximum principal curvatures at each edge midpoint. c,
Direction field showing the directions of minimum principal curvature at the midpoints of edges.
}
  \label{fig:102}
\end{figure}

A different tetriamond is shown in
figure \ref{fig:102}. The bending is again essentially cylindrical, but this time along the shorter
dimension. Now the minimum-curvature direction field is essentially orthogonal to the upper and lower edges, giving
a shape with bilateral symmetry.

\begin{figure}
  \centerline{\includegraphics[width=5.8in] 
  {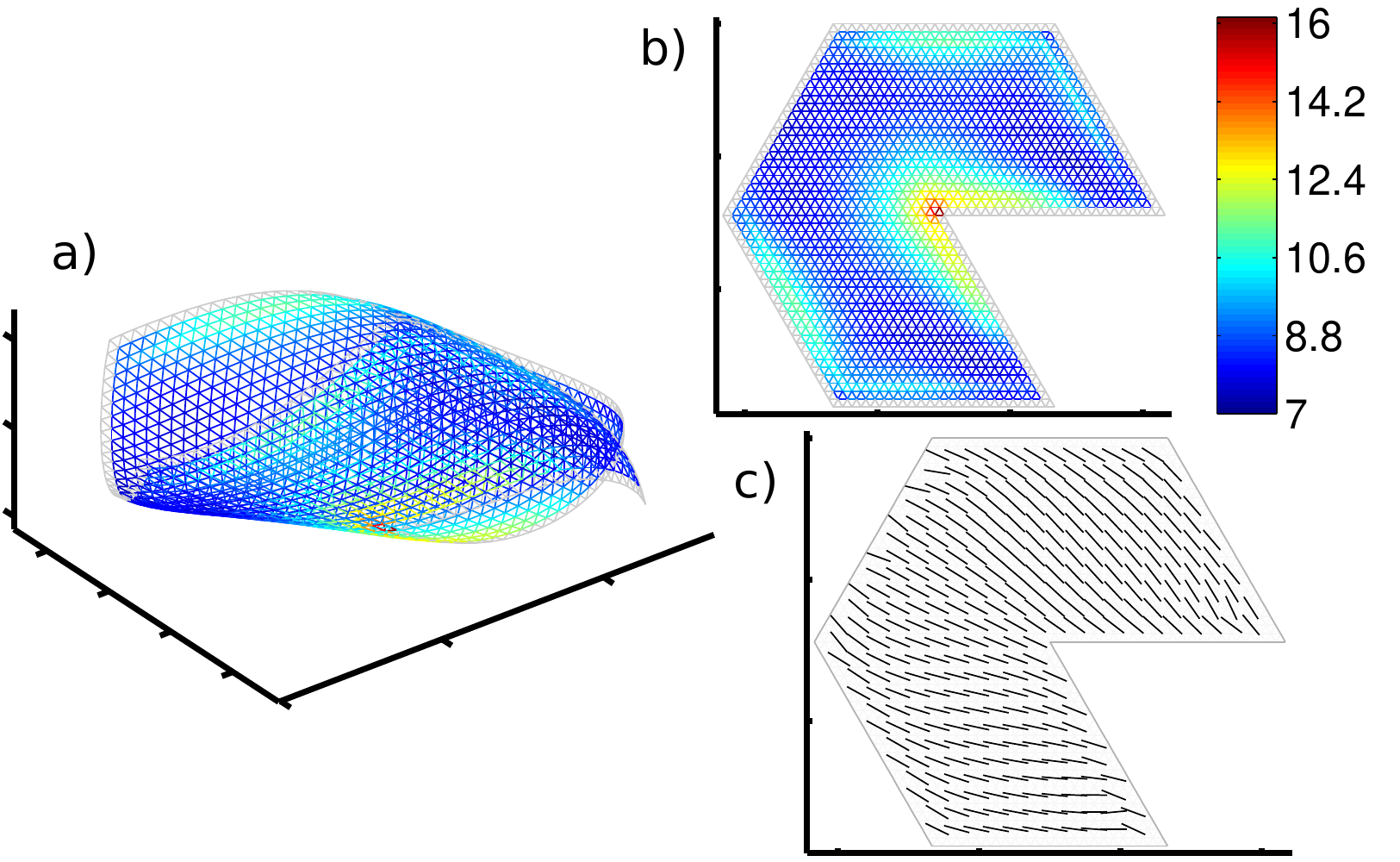}}
\vspace{-0.2in}
  \caption{Bending of a pentiamond bilayer with the same elastic parameters as in
figure \ref{fig:Rect1Fig}. The horizontal width is 1/3. The values on the color bar indicate the maximum of the principal curvatures at each edge midpoint. a, 3D configuration with coloring given in
b, Color map of maximum principal curvatures at each edge midpoint. c,
Direction field showing the directions of minimum principal curvature at the midpoints of edges.
}
  \label{fig:103}
\end{figure}

A pentiamond, or hexagon with a wedge removed, is shown in figure \ref{fig:103}. The 3D configuration
has a bilateral symmetry inherited from the 2D shape. The bending is not quite cylindrical.
The minimum-curvature direction field gradually rotates, moving around the central corner. At the central
corner, the curvature is significantly increased, perhaps indicating a smoothed
version of a conical singularity.

\begin{figure}
  \centerline{\includegraphics[width=5.8in] 
  {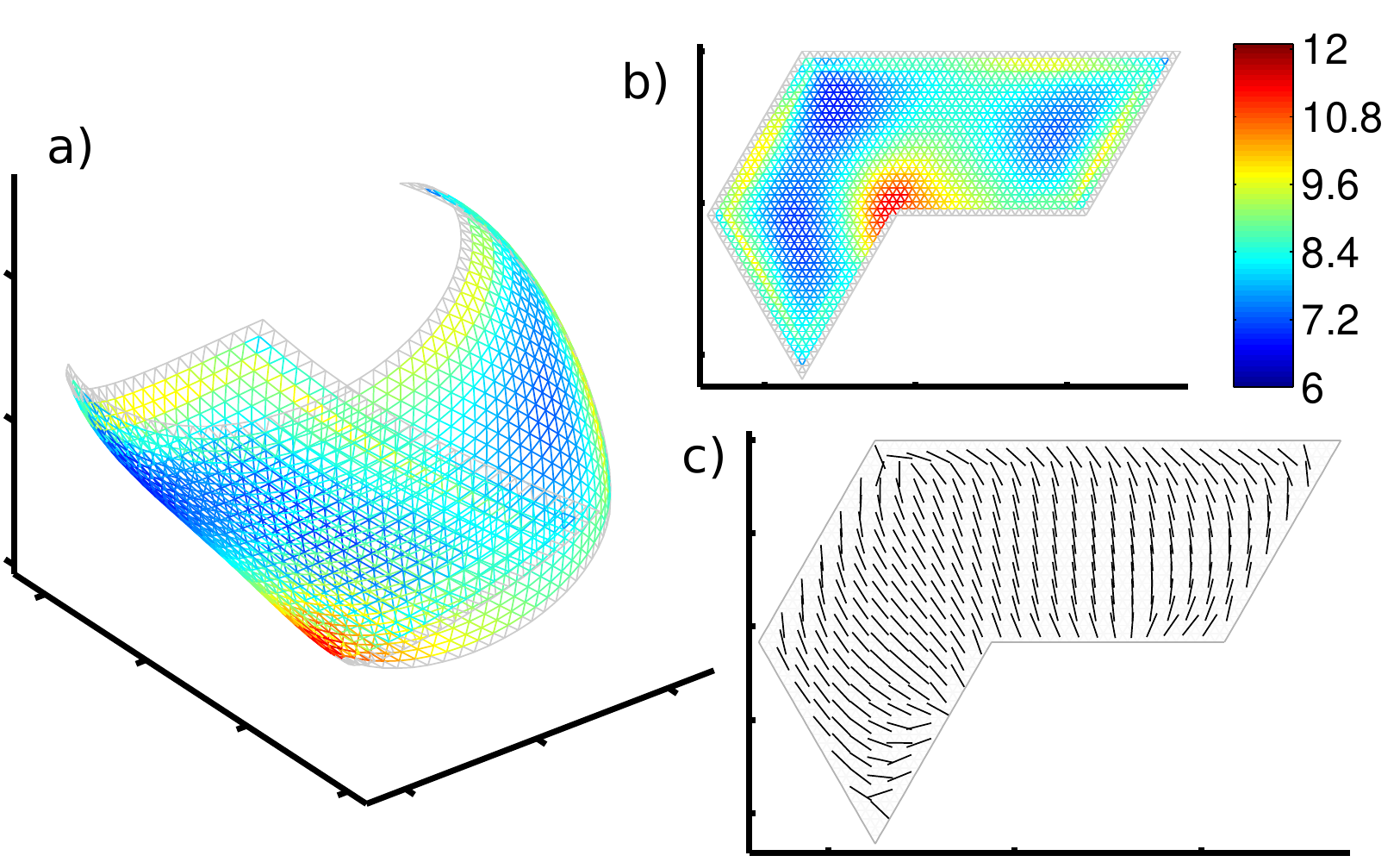}}
\vspace{-0.2in}
  \caption{Bending of a pentiamond bilayer with the same elastic parameters as in
figure \ref{fig:Rect1Fig}. The top edge has length 1/2. The values on the color bar indicate the maximum of the principal curvatures at each edge midpoint. a, 3D configuration with coloring given in
b, Color map of maximum principal curvatures at each edge midpoint. c,
Direction field showing the directions of minimum principal curvature at the midpoints of edges.
}
  \label{fig:104}
\end{figure}

Figure \ref{fig:104} shows the bending of a different pentiamond. Again, the curvature is largest
at the concave angle. The minimum-curvature directions are mainly vertical on the right side of panel c, and
transition to an oblique angle on the left side, yielding an approximate partition into
two regions of bending. Overall, the bending is mainly along the longest dimension of the shape.

\begin{figure}
  \centerline{\includegraphics[width=5.8in] 
  {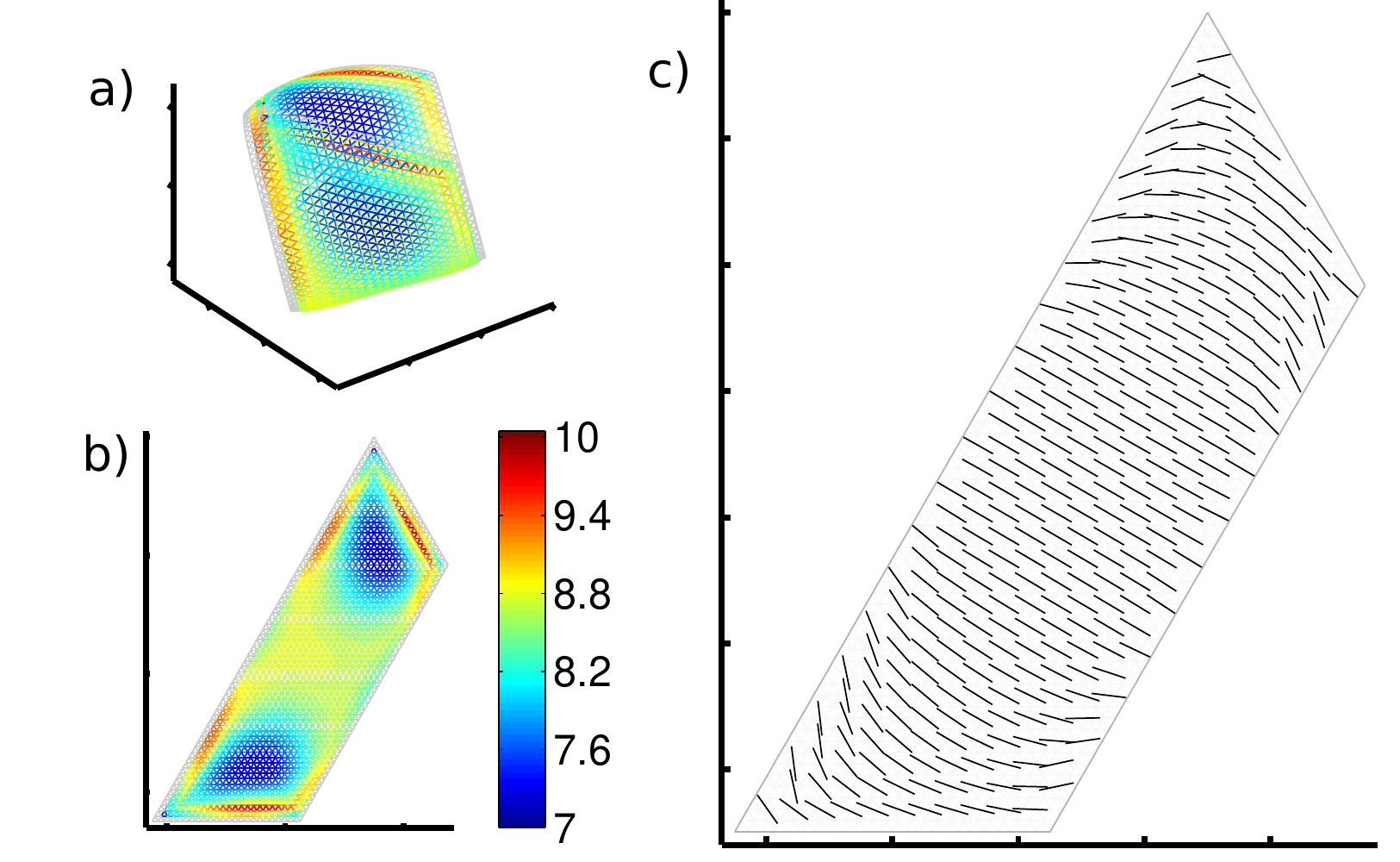}}
\vspace{-0.2in}
  \caption{Bending of a pentiamond bilayer with the same elastic parameters as in
figure \ref{fig:Rect1Fig}. The bottom edge has length 1/4. The values on the color bar indicate the maximum of the principal curvatures at each edge midpoint. a, 3D configuration with coloring given in
b, Color map of maximum principal curvatures at each edge midpoint. c,
Direction field showing the directions of minimum principal curvature at the midpoints of edges.
}
  \label{fig:105}
\end{figure}

A strip-like pentiamond is shown in figure \ref{fig:105}, similar to the
tetriamond of figure \ref{fig:101}. Unlike that shape, here the initial and final
shapes have bilateral symmetry. Similarly to the shape of figure \ref{fig:101}, the
bending is mainly along the longer dimension, and curvature varies most near the two ends.

\begin{figure}
  \centerline{\includegraphics[width=5.8in] 
  {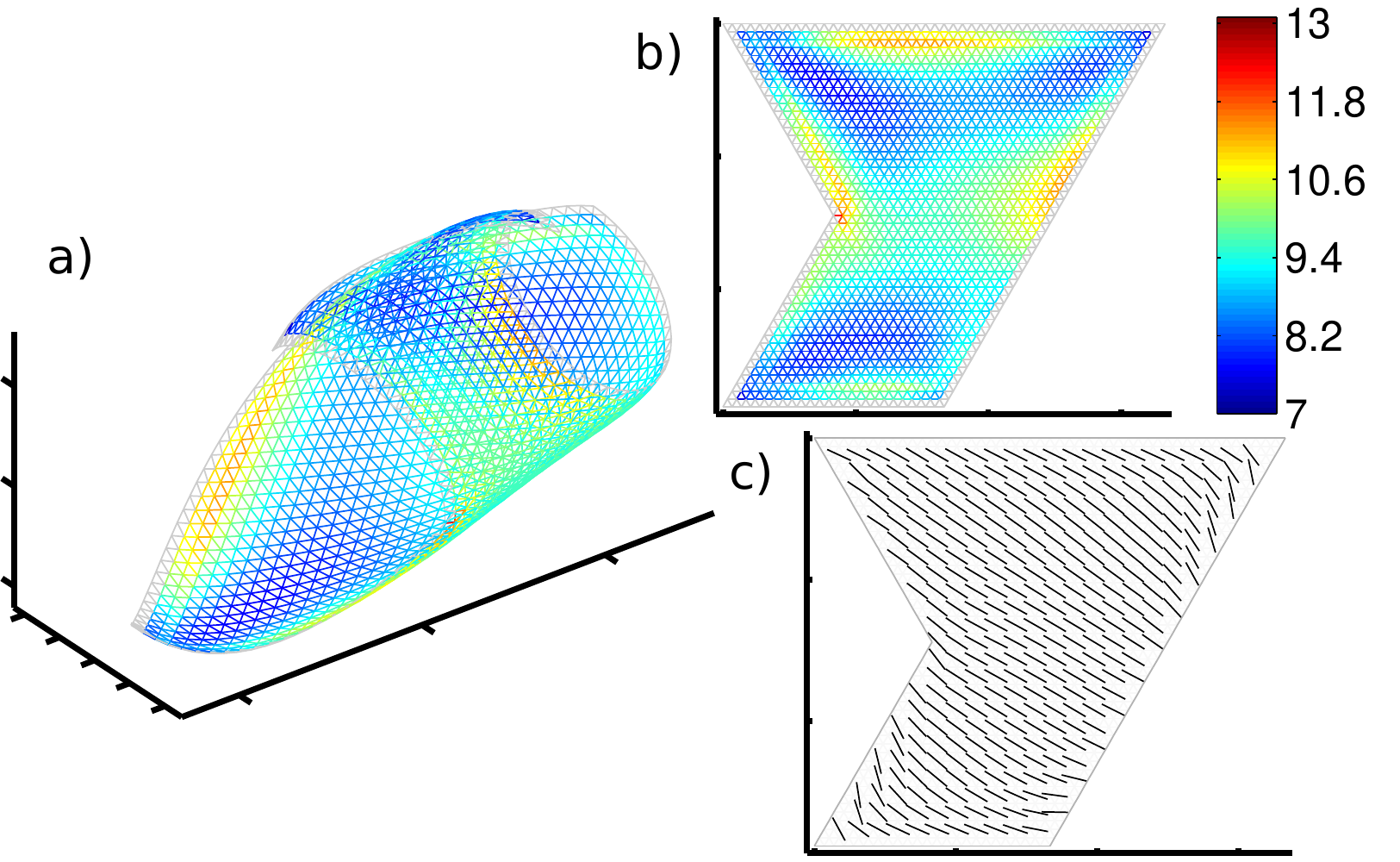}}
\vspace{-0.2in}
  \caption{Bending of a pentiamond bilayer with the same elastic parameters as in
figure \ref{fig:Rect1Fig}. The top edge has length 2/3. The values on the color bar indicate the maximum of the principal curvatures at each edge midpoint. a, 3D configuration with coloring given in
b, Color map of maximum principal curvatures at each edge midpoint. c,
Direction field showing the directions of minimum principal curvature at the midpoints of edges.
}
  \label{fig:107}
\end{figure}

In figure \ref{fig:107}, another pentiamond bilayer is shown. The shape has no symmetries,
and the overall bending is strip-like, with the minimum-curvature directions 
mainly orthogonal to the longest linear dimension of the object.

\begin{figure}
  \centerline{\includegraphics[width=5.8in] 
  {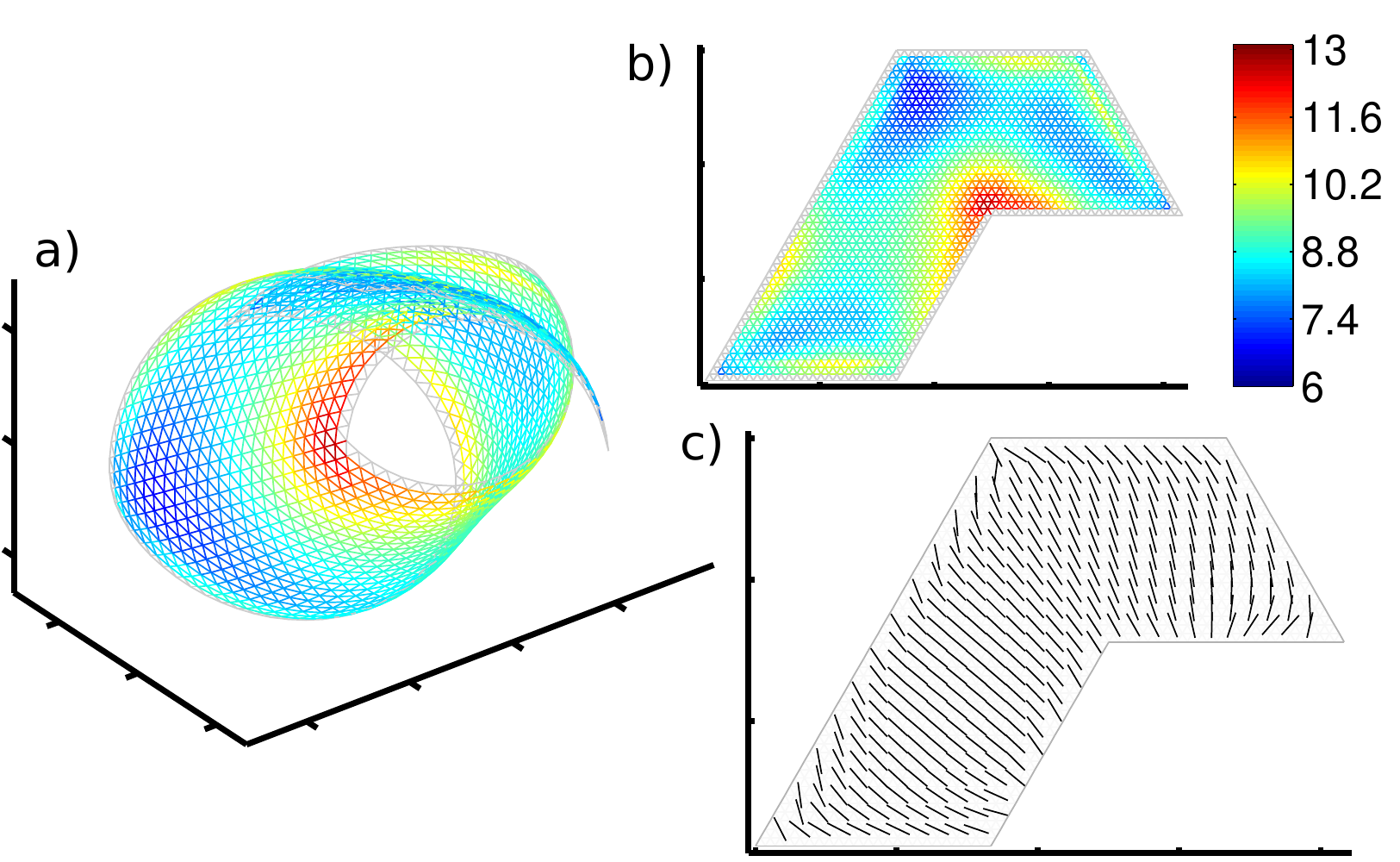}}
\vspace{-0.2in}
  \caption{Bending of a pentiamond bilayer with the same elastic parameters as in
figure \ref{fig:Rect1Fig}. The top edge has length 1/3. The values on the color bar indicate the maximum of the principal curvatures at each edge midpoint. a, 3D configuration with coloring given in
b, Color map of maximum principal curvatures at each edge midpoint. c,
Direction field showing the directions of minimum principal curvature at the midpoints of edges.
}
  \label{fig:108}
\end{figure}

The last of the four pentiamonds is shown in figure \ref{fig:108}, another shape without
symmetries. The bending is most similar to that of figure \ref{fig:104}, with the minimum-curvature direction field
again split mainly into two regions, with some amount of convergence at the concave angle.
The bending is mainly along the longest dimension.

\begin{figure}
  \centerline{\includegraphics[width=5.8in] 
  {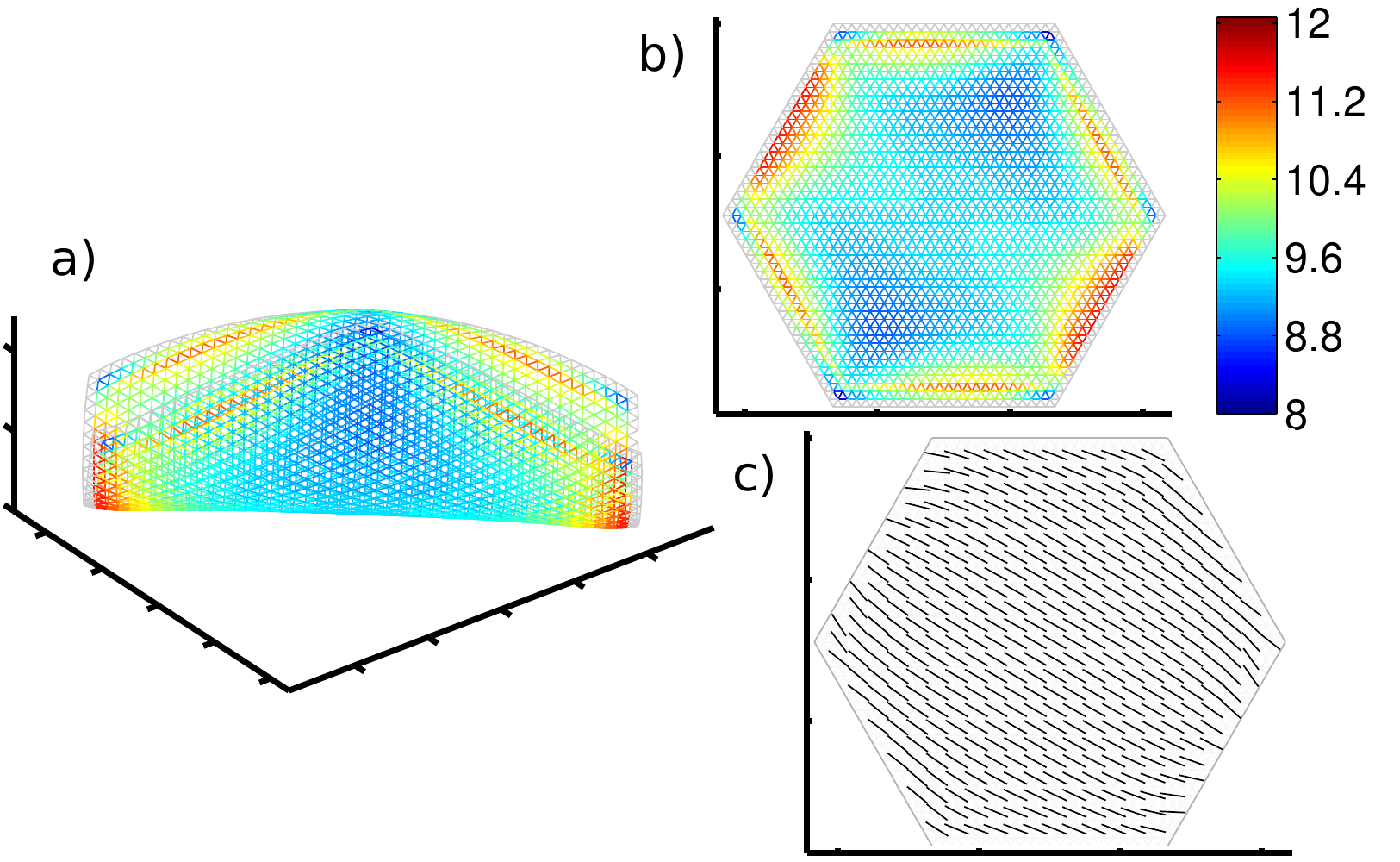}}
\vspace{-0.2in}
  \caption{Bending of a hexagonal bilayer with the same elastic parameters as in
figure \ref{fig:Rect1Fig}. The sides have length 1/3. The values on the color bar indicate the maximum of the principal curvatures at each edge midpoint. a, 3D configuration with coloring given in
b, Color map of maximum principal curvatures at each edge midpoint. c,
Direction field showing the directions of minimum principal curvature at the midpoints of edges.
}
  \label{fig:106}
\end{figure}

We consider only two of the twelve hexiamonds. In figure \ref{fig:106}, a hexagonal
bilayer is shown. The bending is roughly cylindrical, and nearly along a line
connecting two opposite vertices. The curvature distribution has a bilateral symmetry,
with the strongest intensification of curvature at the sides parallel to the bending direction.

\begin{figure}
  \centerline{\includegraphics[width=5.8in] 
  {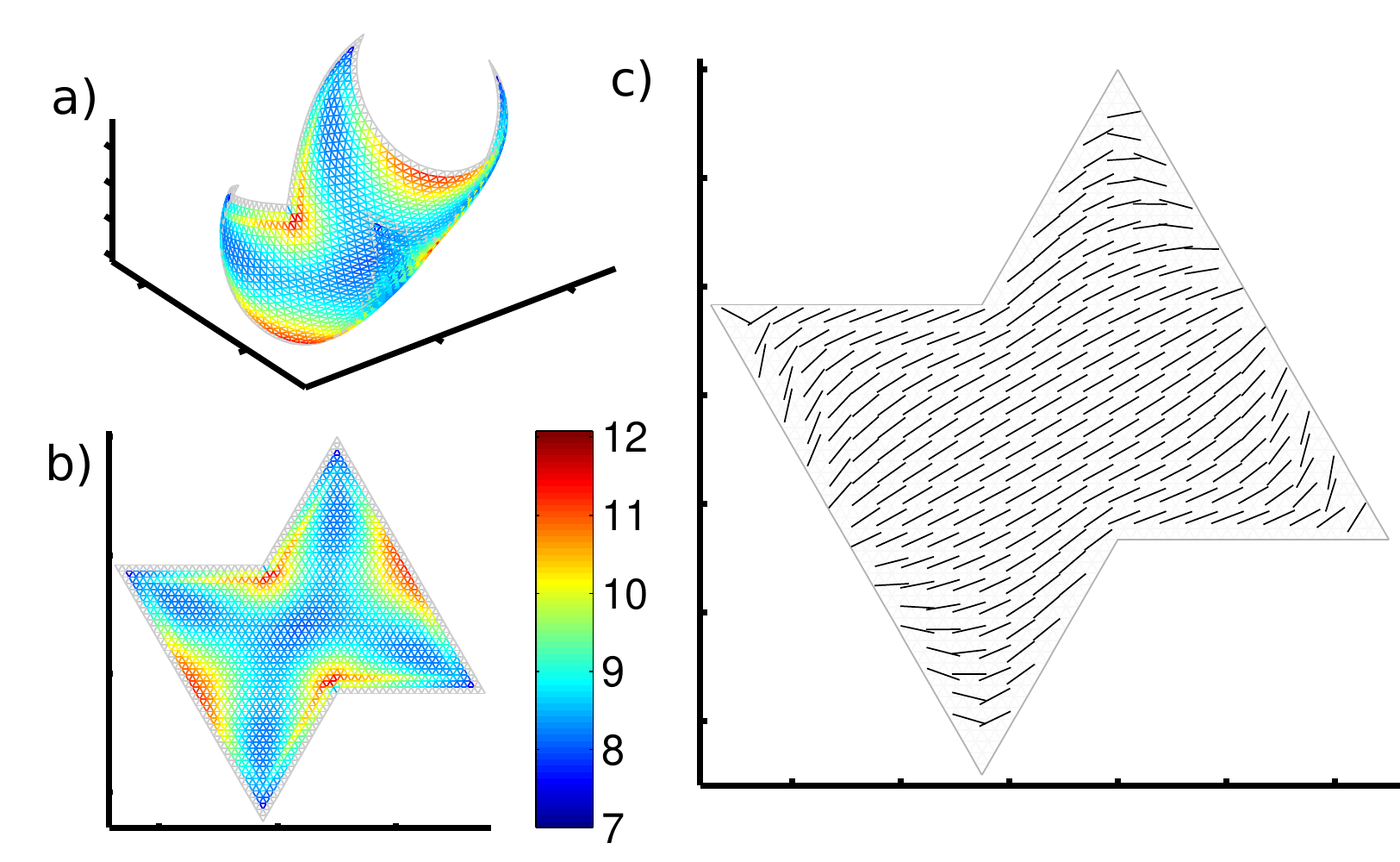}}
\vspace{-0.2in}
  \caption{Bending of a hexiamond with the same elastic parameters as in
figure \ref{fig:Rect1Fig}. The four shorter sides have length 1/3 and the two longer
sides have length 2/3. The values on the color bar indicate the maximum of the principal curvatures at each edge midpoint. a, 3D configuration with coloring given in
b, Color map of maximum principal curvatures at each edge midpoint. c,
Direction field showing the directions of minimum principal curvature at the midpoints of edges.
}
  \label{fig:99}
\end{figure}

A second hexiamond, with rotational and reflectional symmetries, is shown in figure \ref{fig:99}.
The bending is mainly cylindrical here also, but there is a deviation in bending
direction near the 60-degree corners, similar to that which has occured in most but not
all of the 60-degree corners in the bilayers previously discussed. These corners curl inward toward
the center of the shape.

We have presented just a single equilibrium shape for a variety of bilayer shapes,
most corresponding to simple polyiamonds. These shapes show how the simple cylindrical
bending of figures \ref{fig:Rect1Fig} and \ref{fig:Rect3Fig} are modified for oblique
shapes and shapes with different or no symmetries. These examples may be useful test cases
from which to abstract the general dependence of bilayer bending on initial shape.

\section{Conclusion \label{Conclusion}}

We have presented a simple discrete formula for the elastic energy of a bilayer, in terms
of the geometry of a single mesh representing the central surface of the substrate layer.
The formula allows for fast simulations of bending bilayers with general geometries. We have found
good agreement between our simulations of a rectangular bilayer, and an approximate analytical
solution.

Simulations of a set of shapes composed of clusters of equilateral triangles (polyiamonds)
present some generic behaviors of bilayers which may lead to an understanding of the relationship
between bending directions and bilayer shapes. Some of the bilayers show cylindrical bending in the
same direction over the entire bilayer, and often along either the shortest or longest linear
extent of the initial shape. Slight deviations in bending direction occur sometimes but not always
near acute-angled corners of the bilayers, where the shape bends inwards towards the center.
Larger variations in curvature occur near edges and corners, and the curvature is
particularly intensified near obtuse-angled corners, as in the pentiamond which is a
hexagon with a sector removed. In this shape and some others, the direction of bending changes
gradually, leading to bending which is somewhat conical rather than cylindrical. Some other
shapes are nearly partitioned into two regions, with bending along distinct directions in
each region.

Because the bilayers are thin, the equilibria
are all close to developable surfaces. Unlike thin homogeneous sheets, in bilayers there
is always considerable in-plane stretching energy, in the direction of small curvature.
The equilibria we have found preserve many of the initial symmetries of the flat bilayer. Performing our simulations with other initial guesses for the equilibria can produce different and possibly less symmetrical equilibria.

\appendix

\section{Extension due to bending \label{hkappa}}

Let us consider a point $\mathbf X_0$ on the
central surface of the substrate. Let the curvature at $\mathbf X_0$ in a
direction $\mathbf v$ be $\kappa$, and the radius of curvature be $R = 1/\kappa$. Consider the intersection
of the surface with the plane spanned by $\mathbf v$ and the normal to the surface, $\mathbf n$. This
is a curved line passing through $\mathbf X_0$. A short segment of length $l$ near $\mathbf X_0$ is approximated by an arc of a circle tangent to the surface at $\mathbf X_0$ with radius $R$. The set of corresponding points on the
central surface of the actuated layer is a parallel curve, approximated by a concentric arc with radius $R + h$, and length $l + \delta l$. Because the arcs are concentric,
\bq
\frac{l + \delta l}{l} = \frac{R + h}{R} \;\;\mbox{and thus}\;\; \frac{\delta l}{l} = h\kappa,
\eq
\nn so the strain in the actuated layer is $h\kappa$ plus that in the substrate. Furthermore, the ratio
of the curvature of the actuated layer central surface $\kappa_a$ to that of the substrate $\kappa$ is
\bq
\frac{\kappa_a}{\kappa} = \frac{R}{R + h} = \frac{1}{1 + h\kappa} = 1 - h\kappa + O\left((h\kappa)^2\right).
\eq

\section{Discrete formula for extension due to bending \label{dischkappa}}

In order to derive equation (\ref{dhkappa}), we consider the discretization in the vicinity
of a point on the central surface of the substrate, which is assumed to be a smooth surface. Without
loss of generality, we can assume that the point is at the origin in $(x,y,z)$, and
that, up to quadratic order, the central surface of the subtrate is given by
\bq
z(x,y) = \frac{1}{2}k_x x^2 + \frac{1}{2}k_y y^2. \label{quad}
\eq
\nn An arbitrary smooth surface can be brought into this form (up to quadratic order), in the vicinity of a point on the
surface, by translating the point to the origin and rotating the surface about the point so that the directions of principal curvature lie along the $x$ and $y$ axes.

\begin{figure}
  \centerline{\includegraphics[width=5.8in] 
  {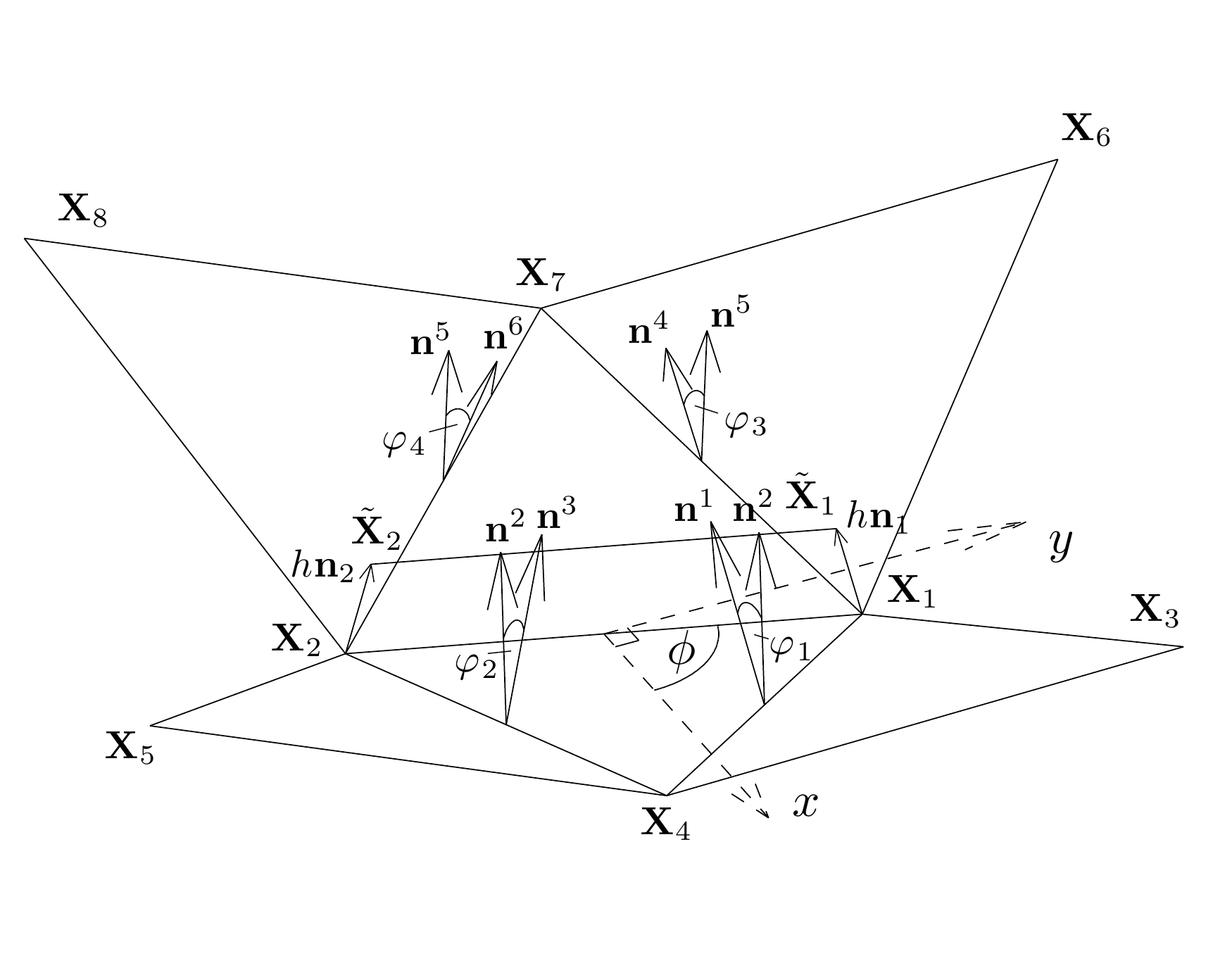}}
\vspace{-0.2in}
  \caption{Schematic diagram of a set of triangles showing the notation used in deriving the formula
for the strain in edge ${\tilde{\mathbf X}}_1$-${\tilde{\mathbf X}}_2$ due to bending. Here the 
amount of bending is exaggerated so the differences between vectors are visible.
}
  \label{fig:PolymerMeshSchematic2}
\end{figure}

We now assume that the origin coincides with the midpoint of an edge in our triangulation of
the substrate central surface. We determine the stretching of this edge in the actuated
layer relative to that in the substrate, under the bending of the substrate given by (\ref{quad}).
In figure \ref{fig:PolymerMeshSchematic2} we show a portion of the triangular mesh consisting of the two triangles which
bound the ${\mathbf X}_1$-${\mathbf X}_2$ edge, and the four other triangles which share an edge with these triangles. We can approximate the curvatures of the surface simply in terms of the angles
between the normals to these triangles. In general the mesh has an arbitrary orientation
with respect to the directions of principal curvature (here, the $x$ and $y$ axes).
Thus in figure \ref{fig:PolymerMeshSchematic2} we assume an arbitrary angle $\phi$ between the $x$-axis and the projection of
the ${\mathbf X}_1$-${\mathbf X}_2$ edge on the $x$-$y$ plane. We first assume that the $x$-$y$ coordinates of the eight points in figure \ref{fig:PolymerMeshSchematic2} lie on the undeformed equilateral triangular lattice, which is a good approximation
for small strains:
\begin{align}
{\mathbf X}_{1,x} &= \frac{d_{eq}}{2}\cos{\phi}, {\mathbf X}_{1,y} = \frac{d_{eq}}{2}\sin{\phi},\label{x1}\\
\left({\mathbf X}_{2,x}, {\mathbf X}_{2,y}\right) &= -\left({\mathbf X}_{1,x}, {\mathbf X}_{1,y}\right)\\
\left({\mathbf X}_{3,x}, {\mathbf X}_{3,y}\right) &= \left({\mathbf X}_{1,x}, {\mathbf X}_{1,y}\right) + d_{eq}\left(\cos{\left(\phi-\frac{\pi}{3}\right)}, \sin{\left(\phi-\frac{\pi}{3}\right)}\right) \\
\left({\mathbf X}_{4,x}, {\mathbf X}_{4,y}\right) &= \left({\mathbf X}_{3,x}, {\mathbf X}_{3,y}\right) - d_{eq}\left(\cos{\phi}, \sin{\phi}\right) \\
\left({\mathbf X}_{5,x}, {\mathbf X}_{5,y}\right) &= \left({\mathbf X}_{4,x}, {\mathbf X}_{4,y}\right) - d_{eq}\left(\cos{\phi}, \sin{\phi}\right) \\
\left({\mathbf X}_{6,x}, {\mathbf X}_{6,y}\right) &= \left({\mathbf X}_{1,x}, {\mathbf X}_{1,y}\right) + d_{eq}\left(\cos{\left(\phi+\frac{\pi}{3}\right)}, \sin{\left(\phi+\frac{\pi}{3}\right)}\right) \\
\left({\mathbf X}_{7,x}, {\mathbf X}_{7,y}\right) &= \left({\mathbf X}_{6,x}, {\mathbf X}_{6,y}\right) - d_{eq}\left(\cos{\phi}, \sin{\phi}\right) \\
\left({\mathbf X}_{8,x}, {\mathbf X}_{8,y}\right) &= \left({\mathbf X}_{7,x}, {\mathbf X}_{7,y}\right) - d_{eq}\left(\cos{\phi}, \sin{\phi}\right) \label{x8}
\end{align}
\nn The $z$ coordinates of ${\mathbf X}_1$-${\mathbf X}_8$ are found by inserting
the $x$-$y$ coordinates in (\ref{x1})-(\ref{x8}) into (\ref{quad}). The points which
correspond to ${\mathbf X}_1$-${\mathbf X}_2$ in the central surface of the
actuated layer are
\begin{align}
{\tilde{\mathbf X}}_1 &= {\mathbf X}_1 + h {\mathbf n}_1 \\
{\tilde{\mathbf X}}_2 &= {\mathbf X}_2 + h {\mathbf n}_2
\end{align}
\nn where ${\mathbf n}_1$ and ${\mathbf n}_2$ are the unit normal vectors to the
surface (\ref{quad}) at ${\mathbf X}_1$ and ${\mathbf X}_2$, respectively:
\begin{align}
{\mathbf n}_1 &= \left(-k_x \frac{d_{eq}}{2}\cos{\phi},\,-k_y \frac{d_{eq}}{2}\sin{\phi},\,1\right)\left( 1+O\left(d_{eq}^2\right)\right)  \\
{\mathbf n}_2 &= \left(k_x \frac{d_{eq}}{2}\cos{\phi},\,k_y \frac{d_{eq}}{2}\sin{\phi},\,1\right)\left( 1+O\left(d_{eq}^2\right)\right).
\end{align}
\nn We may now determine the length of the edge
${\tilde{\mathbf X}}_1$-${\tilde{\mathbf X}}_2$
relative to $d_{eq}$:
\bq
\left|{\tilde{\mathbf X}}_1-{\tilde{\mathbf X}}_2 \right| - d_{eq} = -h d_{eq} (k_x \cos^2\phi + k_y \sin^2\phi) \left( 1+O\left(d_{eq}^2\right)\right)\left( 1+O\left(h\right)\right). \label{astretch}
\eq
\nn If the corresponding edge ${\mathbf X}_1$-${\mathbf X}_2$ in the
substrate has a length $d$ different from $d_{eq}$, then
(\ref{astretch}) still holds but with $d$ in place of $d_{eq}$. Thus it
gives the difference in stretching between the actuated layer
central surface and the substrate at corresponding edges.

We now show that (\ref{astretch})
is approximately $-h\sqrt{3}$ times the average of
the angles between
the normals to the four triangles adjacent to the two triangles which share the
${\mathbf X}_1$-${\mathbf X}_2$ edge. We define vectors normal to triangles 1-6
\begin{align}
{\mathbf n}^1 &= ({\mathbf X}_1-{\mathbf X}_3) \times ({\mathbf X}_4-{\mathbf X}_3) \label{n1} \\
{\mathbf n}^2  &= ({\mathbf X}_4-{\mathbf X}_2) \times ({\mathbf X}_2-{\mathbf X}_1) \label{n2}\\
{\mathbf n}^3  &= ({\mathbf X}_4-{\mathbf X}_5) \times ({\mathbf X}_2-{\mathbf X}_5) \\
{\mathbf n}^4  &= ({\mathbf X}_7-{\mathbf X}_6) \times ({\mathbf X}_1-{\mathbf X}_6) \\
{\mathbf n}^5  &= ({\mathbf X}_7-{\mathbf X}_1) \times ({\mathbf X}_2-{\mathbf X}_1) \\
{\mathbf n}^6  &= ({\mathbf X}_7-{\mathbf X}_2) \times ({\mathbf X}_8-{\mathbf X}_2)
\end{align}
\nn The angle between ${\mathbf n}^1$ and ${\mathbf n}^2$ is
\bq
\varphi_1 = \mbox{Arccos}\left(\frac{{\mathbf n}^1\cdot{\mathbf n}^2}
{\|{\mathbf n}^1\|\|{\mathbf n}^2\|}\right) \label{angle12}
\eq
\nn Using (\ref{n1},\ref{n2}) we define a function $a_1$ by
\bq
\frac{4}{3}{\mathbf n}^1\cdot{\mathbf n}^2 =
d_{eq}^4(1 + d_{eq}^2 a_1(k_x,k_y,\phi)),
\eq
\nn $a_1$ has terms which are quadratic in $k_x$ and $k_y$
and include trigonometric functions of $\phi$. We also define $b_1$ and $c_1$
for terms in the denominator of (\ref{angle12}):
\begin{align}
\frac{4}{3}\|{\mathbf n}^1\|^2 &=
d_{eq}^4(1 + d_{eq}^2 b_1(k_x,k_y,\phi)) \\
\frac{4}{3}\|{\mathbf n}^2\|^2 &=
d_{eq}^4(1 + d_{eq}^2 c_1(k_x,k_y,\phi))
\end{align}
\nn Thus
\bq
\frac{{\mathbf n}^1\cdot{\mathbf n}^2}
{\|{\mathbf n}^1\|\|{\mathbf n}^2\|} =
1 + d_{eq}^2\left(a_1 - \frac{b_1}{2} - \frac{c_1}{2}\right)\left( 1+O\left(d_{eq}^2\right)\right).
\eq
\nn We define $a_2$ and $b_2$ in terms of ${\mathbf n}^2$ and ${\mathbf n}^3$:
\begin{align}
\frac{4}{3}{\mathbf n}^3\cdot{\mathbf n}^2 &=
d_{eq}^4(1 + d_{eq}^2 a_2(k_x,k_y,\phi))\\
\frac{4}{3}\|{\mathbf n}^3\|^2 &=
d_{eq}^4(1 + d_{eq}^2 b_2(k_x,k_y,\phi))
\end{align}
\nn The angle between ${\mathbf n}^2$ and ${\mathbf n}^3$ is
\bq
\varphi_2 = \mbox{Arccos}\left(\frac{{\mathbf n}^2\cdot{\mathbf n}^3}
{\|{\mathbf n}^2\|\|{\mathbf n}^3\|}\right) \label{angle23}
\eq
\nn We expand Arccos in a Taylor series
\bq
\mbox{Arccos}(1-x) = \sqrt{2x} + O(x^{3/2}).
\eq
\nn Thus
\bq
\frac{1}{2}(\varphi_1 + \varphi_2) = \frac{1}{\sqrt{2}}d_{eq}
\left(\sqrt{-a_1 + \frac{b_1+c_1}{2}} + \sqrt{-a_2 + \frac{b_2+c_1}{2}} \right)\left( 1+O\left(d_{eq}^2\right)\right)
\label{phiabc}
\eq
\nn Using trigonometric identities one can show that
\bq
\frac{1}{\sqrt{2}}
\left(\sqrt{-a_1 + \frac{b_1+c_1}{2}} + \sqrt{-a_2 + \frac{b_2+c_1}{2}} \right)
= \frac{1}{\sqrt{3}}\left(k_x \cos^2\phi + k_y \sin^2\phi\right). \label{abck}
\eq
\nn Therefore by (\ref{astretch}),
\bq
-h\sqrt{3}\,\frac{\varphi_1 + \varphi_2}{2}\left( 1+O\left(d_{eq}^2\right)\right)\left( 1+O\left(h\right)\right) = \left|{\tilde{\mathbf X}}_1-{\tilde{\mathbf X}}_2 \right| - d_{eq}. \label{actform1}
\eq
\nn A more symmetrical formula is obtained by using
the angles
\begin{align}
\varphi_3 = \mbox{Arccos}\left(\frac{{\mathbf n}^4\cdot{\mathbf n}^5}
{\|{\mathbf n}^4\|\|{\mathbf n}^5\|}\right) ;\quad
\varphi_4 = \mbox{Arccos}\left(\frac{{\mathbf n}^5\cdot{\mathbf n}^6}
{\|{\mathbf n}^5\|\|{\mathbf n}^6\|}\right) \label{angle4556}
\end{align}
\nn By symmetries of the lattice (\ref{x1})-(\ref{x8}) and the quadratic surface,
$\varphi_3 = \varphi_2$ and $\varphi_4 = \varphi_1$. Thus a
more symmetrical alternative to (\ref{actform1}) is
\bq
-h\sqrt{3}\,\frac{\varphi_1 + \varphi_2 + \varphi_3 + \varphi_4}{4}\left( 1+O\left(d_{eq}^2\right)\right)\left( 1+O\left(h\right)\right) = \left|{\tilde{\mathbf X}}_1-{\tilde{\mathbf X}}_2 \right| - d_{eq}. \label{actform2}
\eq
\nn (\ref{actform2}) is more accurate than (\ref{actform1}) for surface approximations which are of
higher than quadratic order, in which case the principal curvatures have a nonzero gradient at the origin.

Let us now assume that the substrate central surface is strained
from the undeformed triangular mesh, given by (\ref{x1})-(\ref{x8}),
with a nonzero in-plane strain tensor $\alpha_{ij}$. The
derivation leading to (\ref{actform2}) can be repeated, but now there is an additional factor of
$1 + O(\|\alpha\|)$ multiplying the left side of (\ref{actform2}). This error is of the same order
as errors already present in the single-plate model represented by (\ref{Es}) and (\ref{Eb}) alone. These
errors are due to the use of only the linear part of the strain tensor in infinitessimal strain theory, and also
due to the discrete approximation (\ref{Es}) (see \cite{Seung:1988}).

\section{Estimate of principal curvatures at a point \label{pkappas}}

We now use the framework from appendix \ref{dischkappa} to obtain a discrete estimate for the
principal curvatures at the midpoint of an edge in our mesh. We consider the edge
shown in figure \ref{fig:PolymerMeshSchematic2}, oriented at an angle $\phi$ with respect to the $x$ axis,
and $\pi/2 - \phi$ with respect to the $y$ axis. The $x$ and $y$ axes are again assumed
to be the directions of principal curvatures, and near the edge midpoint the
surface is (\ref{quad}). We can use (\ref{actform2}) to estimate the curvature $k_1$ in the direction of the edge ${\mathbf X}_1$-${\mathbf X}_2$ (i.e. along the angle $\phi$ in figure \ref{fig:PolymerMeshSchematic2}):
\bq
h k_1 d_{eq} \approx \frac{\varphi_1 + \varphi_2 + \varphi_3 + \varphi_4}{4}. \label{k1}
\eq
\nn We can similarly estimate the curvatures along the directions $\phi \pm \pi/3$.
Let $k_2$ denote the curvature in the direction $\phi + \pi/3$,
at the midpoint of ${\mathbf X}_1$-${\mathbf X}_2$. We
 apply formula (\ref{k1}) to the edges
${\mathbf X}_2$-${\mathbf X}_7$ and ${\mathbf X}_1$-${\mathbf X}_4$ which
have angle $\phi + \pi/3$, and take the average as an estimate of
the curvature at the midpoint of ${\mathbf X}_1$-${\mathbf X}_2$. We
take the average of (\ref{k1}) applied to the edges
${\mathbf X}_1$-${\mathbf X}_7$ and ${\mathbf X}_2$-${\mathbf X}_4$
to estimate $k_3$, the curvature in direction $\phi - \pi/3$.

The curvatures $k_1$-$k_3$ are related to the principal curvatures $k_x$ and $k_y$ by
\begin{align}
k_1 &= k_x \cos^2 \phi + k_y \sin^2 \phi.\label{k1a}\\
k_2 &= k_x \cos^2\left(\phi + \frac{\pi}{3}\right) + k_y \sin^2\left(\phi + \frac{\pi}{3}\right).\\
k_3 &= k_x \cos^2\left(\phi - \frac{\pi}{3}\right) + k_y \sin^2\left(\phi - \frac{\pi}{3}\right). \label{k3a}
\end{align}
\nn Given our estimates of $k_1$-$k_3$ at the midpoint of a given edge, we regard (\ref{k1a})-(\ref{k3a})
as a system of three nonlinear equations to be solved for three unknowns: the principal curvatures
$k_x$ and $k_y$, and the angle $\phi$ which orients the mesh with respect to the directions of
principal curvature. The equations are simpler in terms of the alternate set of three variables
\bq
\{k_y, u, A\}, \quad u = \cos^2\phi, \quad A = k_x - k_y.
\eq
\nn The solutions are
\begin{align}
u &= \frac{1}{2} +
\frac{(2k_1 - k_2 - k_3)\,\mbox{sign}(k_2-k_3)}{2\sqrt{3(k_3 - k_2)^2 + (2k_1 - k_2 - k_3)^2}}, \quad
A = \frac{k_2-k_3}{\sqrt{3u(1-u)}}, \quad
k_y = k_1 - A u. \label{ksol}
\end{align}
\nn Using (\ref{ksol}) and (\ref{k1a})-(\ref{k3a}) we can estimate the principal
curvatures at the midpoints of all the edges in our mesh except those
near the boundary, for which not all the edges needed to estimate $k_1$,
$k_2$, and $k_3$ exist.

\bibliographystyle{unsrt}
\bibliography{BilayerStructures}

\end{document}